\pdfoutput=1
\newcommand*{\ATLASLATEXPATH}{latex/}
\documentclass[cernpreprint,texlive=2016,txfonts,UKenglish]{latex/atlasdoc}

\usepackage[csquotes=false,default,subfigure]{\ATLASLATEXPATH atlaspackage}
\usepackage{setspace}
\usepackage{rotating}

\usepackage{\ATLASLATEXPATH atlasbiblatex}

\usepackage{\ATLASLATEXPATH atlascontribute}

\usepackage{\ATLASLATEXPATH atlasphysics}

\graphicspath{{logos/}{figures/}}

\usepackage{Moriond2017_HZZllll_Differential-defs}

\usepackage{dcolumn}
\newcolumntype{d}[1]{D{-}{\mbox{--}}{#1}}

\AtlasTitle{Measurement of inclusive and differential cross sections in the $H \rightarrow ZZ^* \rightarrow 4\ell$ decay channel in $pp$ collisions at $\sqrt{s}$ = 13 \tev\ with the ATLAS detector}

\author{The ATLAS Collaboration}

\AtlasRefCode{HIGG-2016-25}

\PreprintIdNumber{CERN-EP-2017-139}

\arXivId{1708.02810}

\AtlasJournalRef{J. High Energ. Phys. (2017) 2017: 132} 
\AtlasDOI{10.1007/JHEP10(2017)132}

\AtlasAbstract{%
Inclusive and differential fiducial cross sections of Higgs boson production in proton--proton collisions are measured in the $H \rightarrow ZZ^* \rightarrow 4\ell$ decay channel.
The proton--proton collision data were produced at the Large Hadron Collider at a centre-of-mass energy of 13~\tev\ and recorded by the ATLAS detector in 2015 and 2016, corresponding to an integrated luminosity of 36.1~\ifb. 
  The inclusive fiducial cross section in the $H \rightarrow ZZ^* \rightarrow 4\ell$ decay channel is measured to be \mxsec\mxsecErr\ fb, in agreement with the Standard Model prediction of \mxsecExp\mxsecExpErr\ fb. The cross section is also extrapolated to the total phase space including all Standard Model Higgs boson decays. 
Several differential fiducial cross sections are measured for observables sensitive to the Higgs boson production and decay, including kinematic distributions of jets produced in association with the Higgs boson. Good agreement is found between data and Standard Model predictions. The results are used to put constraints on  anomalous Higgs boson interactions with Standard Model particles, using the pseudo-observable extension to the kappa-framework.

}

\hypersetup{pdftitle={ATLAS document},pdfauthor={The ATLAS Collaboration}}

\begin{document}

\maketitle
\tableofcontents

\section{Introduction}
\label{sec:intro}
The ATLAS and CMS Collaborations at the Large Hadron Collider (LHC) have performed extensive studies of the Higgs boson properties in the past few years. The Higgs boson mass has been measured to be $\mH = 125.09 \pm 0.24$~\gev~\cite{mass} and no significant deviations from Standard Model (SM) predictions have been found in the cross sections measured per production mode, the branching ratios~\cite{CombCouplings}, or spin and parity quantum numbers~\cite{spin1,spin2,Khachatryan:2014kca,Khachatryan:2016tnr}. Furthermore, inclusive and differential fiducial cross sections of Higgs boson production, defined as background-subtracted event yields corrected for the detector response,  have been measured in proton--proton ($pp$) collisions at a centre-of-mass energy of $\sqrt{s} =$~8~\tev, using the $4\ell$ ($\ell = e, \mu$), $\gamma\gamma$, and $e\nu\mu\nu$ final states  \cite{Aad:2014tca,Aad:2014lwa,Aad:2015lha,Aad:2016lvc,Khachatryan:2015yvw,CMS-HIG-14-016}. The measured differential cross sections are also in good agreement with the SM predictions.

This paper presents a measurement of inclusive and differential fiducial cross sections in the \mbox{\hfourl} decay channel using $pp$ collisions at $\sqrt{s} =$~13~\tev\ recorded with the ATLAS detector. The combined effect of a higher centre-of-mass energy and an integrated luminosity of 36.1~\ifb\ is expected to increase the number of Higgs boson events by a factor of almost four compared to the previous analysis at $\sqrt{s} = $~8~\tev. Significantly larger gains are expected in the regions of the differential distributions that probe higher momentum scales due to increased parton--parton luminosities. The differential cross sections presented in this paper are measured in a fiducial phase space to avoid model-dependent extrapolations.
The observed distributions are corrected for detector inefficiency and resolution.

Fiducial cross sections are presented both inclusively and separately for each of the final states of the \hfourl\ decay ($4\mu$, $2e2\mu$, $2\mu2e$, $4e$). 
Differential fiducial cross sections are presented for various observables that describe Higgs boson production and decay in $pp$ collisions. They are inclusive in the different final states and Higgs boson production mechanisms, such as gluon--gluon fusion (ggF) or vector-boson fusion (VBF).
The Higgs boson transverse momentum\footnote{ATLAS uses a right-handed coordinate system with its origin at the nominal interaction point (IP) in the centre of the detector and the $z$-axis along the beam pipe. The $x$-axis points from the IP to the centre of the LHC ring, and the $y$-axis points upward. Cylindrical coordinates $(r,\phi)$ are used in the transverse plane, $\phi$ being the azimuthal angle around the $z$-axis. The pseudorapidity is defined in terms of the polar angle $\theta$ as $\eta=-\ln\tan(\theta/2)$.} \ptH\ can be used to test perturbative QCD calculations, especially when separated into exclusive jet multiplicities. 
This variable is also sensitive to the Lagrangian structure of the Higgs boson interactions~\cite{Grazzini:2016paz}.
The Higgs boson rapidity distribution \yH\ is sensitive to the parton distribution functions (PDFs) of the colliding protons. The decay variables \costhetastar\ and \mthreefour\ test the spin and parity of the Higgs boson. The variable \costhetastar\ is defined as the magnitude of the cosine of the decay angle of the leading lepton pair in the four-lepton rest frame with respect to the beam axis. The variables \monetwo\ and \mthreefour\ refer to the invariant masses of the leading and subleading lepton pairs and correspond to the invariant masses of the on-shell and off-shell $Z$ bosons produced in the Higgs boson decay.  The number of jets \njets\ produced in association with the Higgs boson and the transverse momentum \jetpt\ of the leading jet  both provide sensitivity to the theoretical modelling of high-\pt\ quark and gluon emission. The invariant mass \mjj\ of the two leading jets in the event is sensitive to different production mechanisms. 
The signed angle between the two leading jets in the transverse plane\footnote{\dphijj\ is defined as $\dphijj = \phi_{j1} - \phi_{j2}$, if $\eta_{j1}>\eta_{j2}$, otherwise $\dphijj = \phi_{j2} - \phi_{j1}$, where $j1$ is the leading and $j2$ the subleading jet.}  \dphijj\ is another observable that tests the spin and parity of the Higgs boson~\cite{Campbell:2006xx}.

Providing fiducial cross sections simplifies the testing of theoretical models with \hfourl\ final states since the response of the detector has been corrected for.  
As an example, the cross section in the \monetwo\ vs \mthreefour\ observable plane is interpreted in the framework of pseudo-observables~\cite{Gonzalez-Alonso:2014eva},  which are derived from on-shell decay amplitudes and provide a generalization of the kappa-framework~\cite{Heinemeyer:2013tqa}. Limits are set on parameters describing anomalous Higgs boson interactions with leptons and $Z$ bosons.

\section{ATLAS detector}
The ATLAS detector~\cite{Aad:2008zzm} is a multi-purpose detector with a forward-backward symmetric cylindrical geometry.
At small radii, the inner detector (ID), immersed in a 2~T magnetic field produced by a thin superconducting solenoid located in front of the calorimeter, is made up of a fine-granularity pixel detector, including the newly installed insertable B-layer~\cite{ATLAS-TDR-19}, a microstrip detector, as well as a straw-tube tracking detector. The silicon-based detectors cover the pseudorapidity range $|\eta| < $~2.5. The gas-filled straw-tube transition radiation tracker complements the silicon tracker at larger radii up to $|\eta| < $~2 and also provides electron identification capabilities based on transition radiation. The electromagnetic (EM) calorimeter is a lead/liquid-argon sampling calorimeter with accordion geometry. The calorimeter is divided into a barrel section covering $|\eta| < $~1.475 and two end-cap sections covering 1.375~$< |\eta| <$~3.2. For $|\eta| < $~2.5 it is divided into three layers in depth, which are finely segmented in $\eta$ and $\phi$. A
thin presampler layer, covering $|\eta| < $~1.8, is used to correct for fluctuations in upstream energy losses.
A hadronic calorimeter in the region $|\eta| < $~1.7 uses steel absorbers and scintillator tiles as the active medium.
A liquid-argon calorimeter with copper absorbers is used in the hadronic end-cap calorimeters, which covers
the region 1.5~$< |\eta| <$~3.2. A forward calorimeter using copper or tungsten absorbers with liquid argon
completes the calorimeter coverage up to $|\eta| =$~4.9. The muon spectrometer (MS) measures the deflection of
muon trajectories within $|\eta| < $~2.7, using three layers of precision drift tube chambers, with cathode strip chambers in
the innermost layer for $|\eta| >$~2.0. The deflection is provided by a toroidal magnetic field from air-core
superconducting magnets.
The field integral of the 
toroids ranges between 2.0 and 6.0 T$\cdot$m across most of the detector.
The muon spectrometer is instrumented with trigger chambers covering $|\eta| < $~2.4.
Events are selected using a first-level trigger implemented in custom electronics, which reduces the event
rate to a maximum of 100~kHz using a subset of detector information. Software algorithms with access
to the full detector information are then used in the high-level trigger to yield a recorded event rate of
about 1~kHz~\cite{Aaboud:2016leb}.

\section{Theoretical predictions and event simulation}
\label{sec:samples}

The Higgs boson production cross sections and decay branching ratios, as well as their uncertainties, are taken from Refs.~\cite{Dittmaier:2011ti,Dittmaier:2012vm,Heinemeyer:2013tqa,deFlorian:2016spz}, and are referred to as LHCXSWG. The cross section for Higgs boson production via ggF is available at next-to-next-to-next-to-leading order (N3LO) in QCD and has next-to-leading-order (NLO) electroweak (EW) corrections applied~\cite{Djouadi:1991tka,Dawson:1990zj,Spira:1995rr,Harlander:2002wh,Anastasiou:2002yz,Ravindran:2003um,Anastasiou:2014vaa,Anastasiou:2014lda,Anastasiou:2016cez,Aglietti:2004nj,Actis:2008ug,deFlorian:2012yg,Anastasiou:2012hx,Baglio:2010ae}.
The cross section for the VBF process is calculated with full NLO QCD and EW corrections~\cite{Ciccolini:2007jr,Ciccolini:2007ec,Arnold:2008rz}, and approximate next-to-next-to-leading-order (NNLO) QCD corrections are applied~\cite{Bolzoni:2010xr}.
The cross sections for the production of an electroweak boson in association with a Higgs boson, $VH$ ($V$ = $W,Z$), are calculated at NNLO accuracy in QCD~\cite{Han:1991ia,Brein:2003wg} and NLO EW radiative corrections~\cite{Ciccolini:2003jy} are applied. 
The cross section for the associated production of a Higgs boson with a $t\bar{t}$ pair, $t\bar{t}H$, is calculated at NLO accuracy in QCD~\cite{Beenakker:2001rj,Beenakker:2002nc,Dawson:2002tg,Dawson:2003zu}.
The cross section for the $b\bar{b}H$ process is calculated by the Santander matching of the five-flavour scheme (NNLO in QCD) and four-flavour scheme (NLO in QCD)~\cite{Harlander:2011aa}. The composition of the different production modes in the SM is 87.3\% (ggF), 6.8\% (VBF), 4.1\% ($VH$), 0.9\% (\ttH), 0.9\% (\bbH).

The Higgs boson decay branching ratio to the four-lepton final state ($\ell = e, \mu$) for $m_H =$ 125~\gev\ is predicted to be 0.0124\%~\cite{Djouadi:1997yw} in the SM using \progname{PROPHECY4F} \cite{Bredenstein:2006rh,Bredenstein:2006ha}, 
which includes the complete NLO QCD and EW corrections, and the interference effects between identical final-state fermions.
Due to the latter, the expected branching ratios of the 4$e$ and 4$\mu$ final states are about 10\% higher than the branching ratios to $2e2\mu$ and $2\mu2e$ final states.

The \POWHEG -\progname{Box} v2 Monte Carlo (MC) event generator~\cite{powheg1,powheg2,powheg3} is used to simulate ggF~\cite{Campbell:2012am}, VBF~\cite{powheg5} and $VH$~\cite{Luisoni:2013kna} processes, using the PDF4LHC NLO PDF set~\cite{Butterworth:2015oua}.
The ggF Higgs boson production is accurate to NNLO in QCD, using the  \POWHEG\ method for merging the NLO Higgs boson plus jet cross section with the parton shower, and the \progname{MiNLO} method~\cite{Hamilton:2012np,Hamilton:2012rf} to simultaneously achieve NLO accuracy for inclusive Higgs boson production. Furthermore, a reweighting procedure is performed using the \progname{HNNLO} program~\cite{Catani:2007vq,Grazzini:2008tf,Grazzini:2013mca} to achieve full NNLO accuracy~\cite{Hamilton:2013fea}. This sample is referred to as \progname{NNLOPS}.
The VBF and $VH$ samples are produced at NLO accuracy in QCD. For $VH$, the \progname{MiNLO} method is used to merge zero- and one-jet events.
For Higgs boson production in association with a heavy quark pair, events are simulated at NLO with \progname{MadGraph5\_aMC@NLO} (v.2.2.3 for \ttH\ and v.2.3.3 for \bbH~\cite{Wiesemann:2014ioa,})~\cite{Alwall:2014hca}, using the CT10nlo PDF set~\cite{Lai:2010vv} for \ttH\ and the NNPDF23 PDF set~\cite{Ball:2012cx} for \bbH.
For the ggF, VBF, $VH$, and \bbH\ production mechanisms, \PYTHIA 8~\cite{pythia,Sjostrand:2007gs} is used for the \hfourl\ decay as well as for parton showering, hadronization, and multiple partonic interactions using the AZNLO parameter set~\cite{AZNLO}.
For the \ttH\ production mechanism, \progname{Herwig++}~\cite{Bahr:2008pv,Bellm:2013hwb} is used with the UEEE5 parameter set~\cite{Seymour:2013qka}.

The measured event yields and the differential fiducial cross-section measurements are compared to a SM prediction constructed from the MC predictions presented above, after normalizing each sample using the corresponding LHCXSWG prediction.
All samples are generated with \hbox{$\mH = 125$~\gev}.

An alternative prediction for ggF SM Higgs boson production is generated using \progname{MadGraph5\_aMC@NLO} v.2.3.3 at NLO accuracy in QCD for zero, one, two additional jets, merged with the FxFx scheme~\cite{Alwall:2014hca,Frederix:2012ps}, using the NNPDF30\_nlo\_as\_0118  PDF set~\cite{Ball:2014uwa}. This \mgfxfx\ sample is interfaced to \hbox{\PYTHIA 8} for Higgs boson decay, parton showering, hadronization and multiple partonic interactions using the A14 parameter set~\cite{ATL-PHYS-PUB-2014-021}.
The data are also compared to ggF SM Higgs boson production in the 4$\ell$ decay channel simulated with \progname{HRes} v2.3~\cite{deFlorian:2012mx,Grazzini:2013mca}, using the MSTW2008 NNLO PDF set~\cite{Martin:2009iq}. The \progname{HRes} program computes fixed-order cross sections for ggF SM Higgs boson production up to NNLO in QCD and describes the \ptH\ distribution at NLO. All-order resummation of soft-gluon effects at small transverse momenta is consistently included up to next-to-next-to-leading logarithmic order (NNLL) in QCD, using dynamic factorization and resummation scales (the central scales are chosen to be $m_H/2$). The program implements top quark and bottom quark mass dependence up to next-to-leading logarithmic order (NNL) + NLO in QCD. At NNLL + NNLO accuracy only the top quark contribution is considered. \progname{HRes} does not perform parton showering and QED final-state radiation effects are not included.
Both the \mgfxfx\ and the \progname{HRes} predictions are normalized using the LHCXSWG cross section.

A ggF sample used to study deviations from the SM predictions within the pseudo-observable framework~\cite{Gonzalez-Alonso:2014eva,Gonzalez-Alonso:2015eva} is generated with  \progname{MadGraph5} at LO using FeynRules~2 \cite{Alloul:2013bka} and the NN23PDF PDF set. The sample is interfaced to \PYTHIA 8 using the A14 parameter set. It is normalized using the LHCXSWG cross section.

The $ZZ^{(*)}$ continuum background from quark--antiquark annihilation is simulated with \progname{Sherpa} 2.2~\cite{Gleisberg:2008ta,Gleisberg:2008fv,Cascioli:2011va}, using the NNPDF3.0 NNLO PDF set. NLO accuracy is achieved in the matrix element calculation for zero- and one-jet final states and LO accuracy for two- and three-jet final states. The merging is performed with the \progname{Sherpa} parton shower~\cite{Schumann:2007mg} using the \progname{MePs}@NLO prescription~\cite{Hoeche:2012yf}.
NLO EW corrections are applied as a function of the invariant mass of the $ZZ^*$ system $m_{\zzstar}$~\cite{Biedermann:2016yvs,Biedermann:2016lvg}.
The gluon-induced \zzstar\ production is modelled with gg2VV~\cite{Kauer:2015dma} at leading order in QCD.
The $K$-factor accounting for missing higher-order QCD effects in the calculation of the $gg\rightarrow ZZ^{*}$ continuum is taken to be 1.7 $\pm$ 1.0~\cite{Caola:2015psa,Caola:2015rqy,Campbell:2016ivq,Bonvini:2013jha,Li:2015jva,Alioli:2016xab}.

\progname{Sherpa} 2.2 is also used to generate samples of the \zjets\ background at NLO accuracy for zero-, one- and two-jet final states and LO accuracy for three- and four-jet final states. In this measurement, the \zjets\ background is normalized using control samples from data. For comparisons with simulation, the QCD NNLO \progname{Fewz}~\cite{Melnikov:2006kv,Anastasiou:2003ds} and \progname{Mcfm} cross-section calculations are used for inclusive $Z$ boson and $Z + b\bar{b}$ production, respectively. 
Samples for the $t\bar{t}$ background are produced with \POWHEG -\progname{Box} interfaced to \PYTHIA 6~\cite{pythia} for parton showering and hadronization, to \progname{Photos}~\cite{Golonka:2005pn} for QED radiative corrections, to \progname{Tauola}~\cite{Jadach:1993hs,Golonka:2003xt} for the simulation of $\tau$ lepton decays and to \progname{EvtGen} v.1.2.0~\cite{Lange:2001uf} for the simulation of $b$-hadron decays. For this sample, the Perugia 2012 parameter set~\cite{Skands:2010ak} is used.
The $WZ$ background is modelled using \POWHEG -\progname{Box}+\PYTHIA 8 and the AZNLO parameter set.
The triboson backgrounds $ZZZ$, $WZZ$, and $WWZ$ with four or more leptons originating from the hard scatter are produced with \progname{Sherpa} 2.1.
\progname{MadGraph}, interfaced to \PYTHIA 8 with the A14 parameter set is used to simulate the all-leptonic $t\bar{t}+Z$ as well as the $t\bar{t}+W$ processes.

The particle-level  events produced by each event generator are passed through the \progname{Geant4}~\cite{GEANT4} simulation of the  ATLAS detector~\cite{simuAtlas} and reconstructed in the same way as the data.  Additional $pp$ interactions in the same and nearby bunch crossings (pile-up) are simulated using inelastic $pp$ collisions generated using \PYTHIA 8 (with the A2 MSTW2008LO parameter set) and overlaid on the simulated events discussed above. The MC events are weighted to reproduce the distribution of the average number of interactions per bunch crossing observed in the data.

\section{Event selection}
\label{sec:eventsel}

Events with at least four leptons are selected with single-lepton, dilepton and trilepton triggers.
The trigger selection requirements, e.g. the minimum transverse energy \et/transverse momentum \pt, the identification and the isolation requirements,  were tightened periodically during the data-taking to maintain a maximum overall trigger rate as the instantaneous luminosity increased. For example, the \et\ threshold changed from 24 to 26~\gev\ for  the single-electron trigger. The multilepton triggers have lower \et\ or \pt\ requirements and more relaxed identification requirements.
The combined trigger efficiency in this analysis is about 98\%. The data are subjected to quality requirements to reject events in which detector components were not operating correctly.  Events are required to have at least one vertex with two associated tracks with \pt~$>$~400~\mev, and the primary vertex is chosen to be the reconstructed vertex with the largest $\sum \pt^2$ of reconstructed tracks. 

Electrons are reconstructed using tracks in the ID and energy clusters in the EM calorimeter~\cite{ATLAS-CONF-2016-024}. They are required to satisfy loose identification criteria based on tracking and calorimeter information. Muons are reconstructed as tracks in the ID and the MS~\cite{Aad:2016jkr} if they lie in the region $0.1< |\eta| < 2.5$. In the region $|\eta| < 0.1$, the MS has reduced coverage, and muons are reconstructed from ID tracks and identified by either a minimal energy deposit in the calorimeter or hits in the MS. For $ 2.5 < |\eta| < 2.7$, only the MS can be used. For events with four muons, at least three muons are required to be reconstructed by combining ID and MS tracks. Each muon (electron) must have transverse momentum $\pt > 5$~\gev\ ($\et > 7$~\gev), within the pseudorapidity range $|\eta| < 2.7$ (2.47) and with a longitudinal impact parameter $|z_0 \sin (\theta)| < 0.5$~mm.  Muons originating from cosmic rays are removed with the transverse impact parameter requirement $|d_0| < 1$ mm.
Jets are reconstructed from topological clusters of calorimeter cells using the 
anti-$k_t$ algorithm~\cite{Cacciari:2008gp,Cacciari:2011ma} with the radius parameter $R = 0.4$.  Jets are corrected for detector response and pile-up contamination  ~\cite{Aaboud:2017jcu,Collaboration:2037613}  and required to have $\pt > 30$~\gev, and $|\eta| < 4.5$. In order to avoid double counting of electrons also reconstructed as jets, jets are removed if $\Delta R(\mathrm{jet}, e) = \sqrt{\Delta \phi(\mathrm{jet}, e)^2 + \Delta \eta(\mathrm{jet},e)^2 } < 0.2$. This overlap removal is also applied to jets close to muons if the jet has fewer than three tracks and the energy and momentum differences between the muon and the jet are small ($p_{\mathrm{T},\mu}  > 0.5\ p_\mathrm{T,jet}$ and $p_{\mathrm{T},\mu}  > 0.7\ p_\mathrm{T,jet,tracks}$), or if $\Delta R(\mathrm{jet}, \mu) < 0.1$.

Higgs boson candidates are formed by selecting two same-flavour opposite-sign (SFOS) lepton pairs, called a lepton quadruplet.
The analysis selection proceeds in parallel for the four final states ($4\mu$, $2e2\mu$, $2\mu2e$, $4e$, where the first two leptons refer to the leading lepton pair). The leading pair is defined as the SFOS pair with the mass \monetwo\ closest to the $Z$ boson mass and the subleading pair is defined as the SFOS pair with the mass \mthreefour\ second closest to the $Z$ boson mass. 
Mispairing within a quadruplet occurs for about 1\% of the selected events for the 4$\mu$ or 4$e$ final states. Furthermore, a quadruplet can be formed with an extra lepton originating from the $W/Z$ for $VH$ or \ttH\ production, moving $m_{4\ell}$ away from $m_H$.  The expected rate for $VH$ or \ttH\ with leptonic decays is about 0.3\% of all Higgs events in the full \mfourl\ range after the event selection.
For each final state, a quadruplet is chosen in which the three leading leptons pass \pt\ (\et) $>$ 20, 15, 10~\gev.
In addition to the dilepton  mass, lepton separation and $J/\psi$ veto requirements (given in Table~\ref{tab:FiducialRegion}), loose calorimeter- and track-based isolation as well as impact parameter requirements are imposed on the leptons. For the track-based isolation, the sum of the \pt\ of the tracks lying within a cone of size $\Delta R = \mathrm{min}[0.3, 10~\gev/\pt]$ ($\mathrm{min}[0.2, 10~\gev/\et]$) around the muon (electron) is required to be smaller than 15\% of the lepton \pt\ (\et). Similarly, the sum of the calorimeter \et\ deposits in a cone of size $\Delta R  = 0.2$ around the muon (electron) is required to be smaller than 30\% (20\%) of the lepton \pt\ (\et). As the four leptons should originate from a common vertex, a requirement on the $\chi^2$ value of a common vertex fit is applied, corresponding to a signal efficiency of 99.5\% for all decay channels.
If more than one quadruplet passes all requirements, e.g. for $VH$ or \ttH, the channel
with the highest expected signal rate after reconstruction and event selection is selected, in the order: $4\mu$, $2e2\mu$, $2\mu 2e$ and $4e$.
In order to improve the four-lepton mass reconstruction, the reconstructed final-state radiation (FSR) photons in $Z$ boson decays are accounted for using the same strategy as in the Run-1 data analysis~\cite{Aad:2014eva}.
The invariant mass distribution of the four leptons of the selected events is shown in Figure~\ref{fig:m4lall}. Only events with a four-lepton invariant mass in the
range 115$-$130~\gev\ are used in the extraction of the signal.

\begin{figure}[!htbp]
\centering
\includegraphics[width=0.7\linewidth]{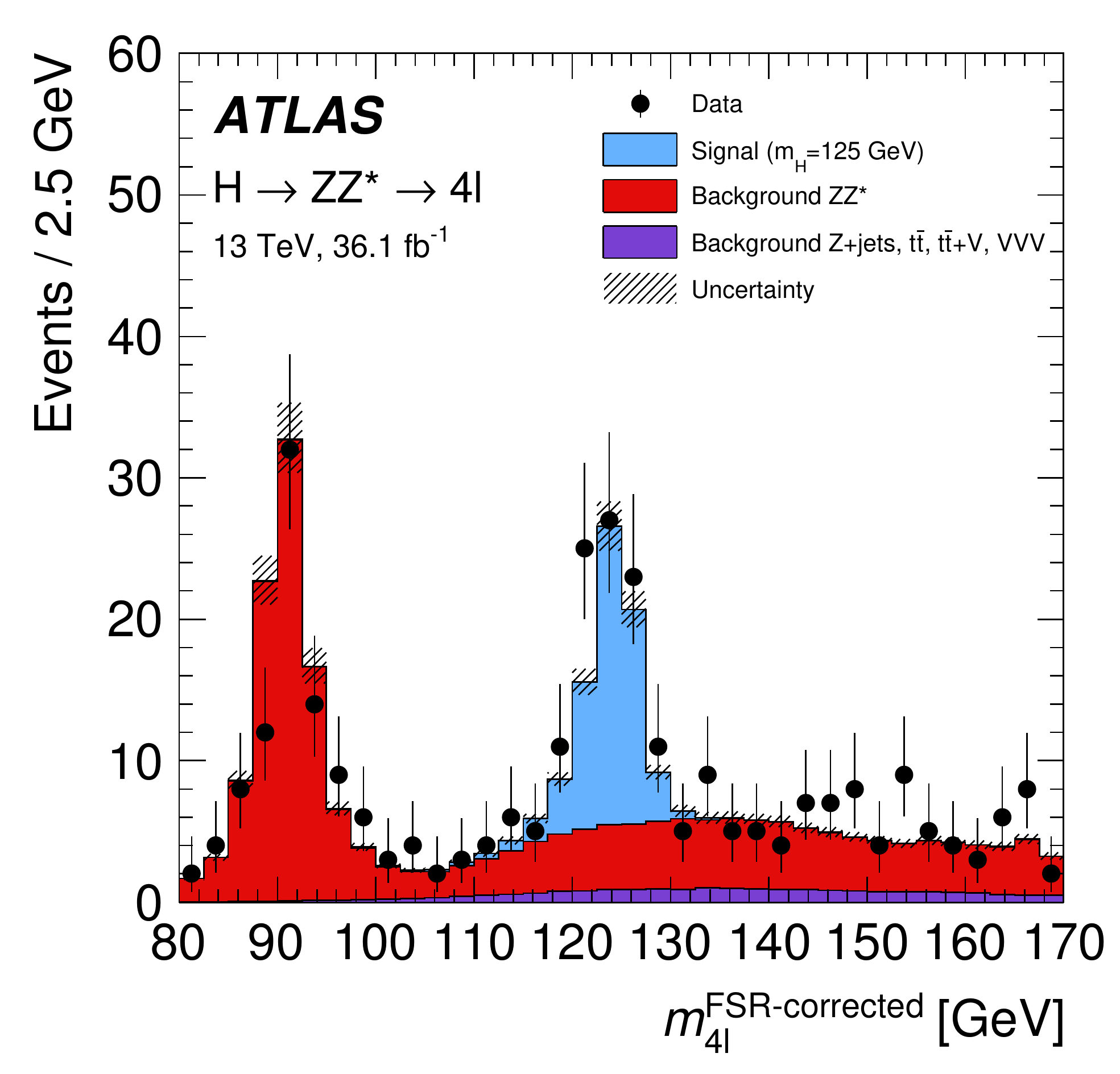}
\caption{Four-lepton invariant mass distribution of the selected events before the \mfourl\ requirement, corrected for final-state radiation (FSR). The error bars on the data points indicate the statistical uncertainty. The SM Higgs boson signal prediction is obtained from the samples discussed in Section~\ref{sec:samples}. The backgrounds are determined following the description in Section~\ref{sec:signalextraction}.  The uncertainty in the prediction is shown by the hatched band, calculated as described in Section~\ref{sec:systematics}. \label{fig:m4lall}}
\end{figure}

The selected events are divided into bins of the variables of interest. 
The bin boundaries are chosen such that each bin has an expected signal significance greater than 2$\sigma$ (where the significance is calculated from the number of signal events $S$ and the number of background events $B$ as $S/\sqrt{S+B}$) and that there are minimal migrations between bins, which reduces the model dependence of the correction for the detector response.

\section{Fiducial phase space}
\label{sec:fiducial}

The fiducial cross sections are defined at particle level using the selection requirements outlined in Table~\ref{tab:FiducialRegion}, which are chosen to closely match those in the detector-level analysis in order to minimize model-dependent acceptance extrapolations. 

The fiducial selection is applied to final-state\footnote{Final-state particles are defined as particles with a lifetime $c\tau > 10$~mm. For electrons and muons, this corresponds to leptons after final state radiation.} electrons and muons that do not originate from hadrons or $\tau$ decays. The leptons are ``dressed'', i.e. the four-momenta of photons within a cone of size $\Delta R$ = 0.1 are added to the lepton four-momentum, requiring the photons to not originate from hadron decays. 
Particle-level jets are reconstructed from final-state particles using the anti-$k_t$ algorithm  with radius parameter $R$~=~0.4. Electrons, muons, neutrinos (if they are not from hadron decays) and photons used to dress leptons, are excluded from the jet clustering. Jets are removed if they are within a cone of size $\Delta R$ = 0.1 (0.2) around a selected muon (electron).

Quadruplets are formed with the selected dressed leptons. Using the same procedure as for reconstructed events reproduces the mispairing of the leptons from Higgs boson decays when assigning them to the leading and subleading $Z$ bosons and the inclusion of leptons originating from vector bosons produced in association with the Higgs boson.
The variables used in the differential cross-section measurement are calculated using the dressed leptons in the quadruplets.

The acceptance of the fiducial selection (with respect to the full phase space of 
$H \rightarrow\ ZZ^* \rightarrow 2\ell 2\ell'$, where $\ell,\ell' = e~\mathrm{or}~\mu$) is 42\% for a SM Higgs boson with $\mH = 125$~\gev. The ratio of the number of events passing the detector-level event selection 
to those passing the particle-level selection is 53\%.
Due to resolution effects, about 2\% of the events which pass the detector-level selection fail the particle-level selection.

\begin{table}
\centering
\caption{List of event selection requirements which define the fiducial phase space of the  cross-section measurement. SFOS lepton pairs are same-flavour opposite-sign lepton pairs. \label{tab:FiducialRegion} }
\vspace{0.2cm}
\begin{tabular}{ll}
\hline\hline
\multicolumn{2}{c}{\textbf{Leptons and jets}} \\
Muons: & $\pt>5$ \GeV, $|\eta|<2.7$  \\
Electrons: & $\pt>7$ \GeV, $|\eta|<2.47$  \\
Jets: &  \pt~$>$ 30 \gev, $|y|<$ \,4.4 \\
Jet--lepton overlap removal: & $\Delta R( \mathrm{jet},\ell) > 0.1\,(0.2)$  for muons (electrons)  \\
\hline
\multicolumn{2}{c}{\textbf{Lepton selection and pairing}} \\
Lepton kinematics: & $\pt > 20, 15, 10 $ \GeV  \\
Leading pair ($m_{12}$): & SFOS lepton pair with smallest $|m_Z-m_{\ell\ell}|$  \\
Subleading pair ($m_{34}$): & remaining SFOS lepton pair with  smallest $|m_Z-m_{\ell\ell}|$  \\
\hline
\multicolumn{2}{c}{\textbf{Event selection (at most one quadruplet per channel)}} \\
Mass requirements: & 50 \GeV$ < m_{12} < 106$ \GeV \, and  12 \GeV $< m_{34} < 115$ \GeV  \\
Lepton separation: & $\Delta R(\ell_{i},\ell_{j}) > 0.1\,(0.2)$  for same- (different-)flavour leptons \\
$J/\psi$~veto: & $m(\ell_{i},\ell_{j}) > 5$ \GeV \,  for all SFOS lepton pairs  \\
Mass window: &  115~\GeV $< m_{4\ell} < 130 $ \GeV  \\
\hline\hline
\end{tabular}

\end{table}

\section{Background estimates}
\label{sec:signalextraction}

Non-resonant SM $ZZ^*$ production via $q\bar{q}$ annihilation and gluon--gluon fusion can result in four prompt leptons in the final state and constitutes the largest background for this analysis. It is estimated using the \progname{Sherpa} and gg2VV simulated samples presented in Section~\ref{sec:samples}. To cross-check the theoretical modelling of this background, a $ZZ^*$-enriched control region is formed using almost the full event selection, but requiring that the four-lepton invariant mass not lie within the region 115 \gev\ $< \mfourl < $130 \gev. In this control region, good agreement is observed between the simulation and the data for all distributions, as demonstrated for \ptH\ and \njets\  in Figure~\ref{fig:ZZCR}.

\begin{figure}[!htbp]
\centering
\subfigure[\label{fig:cr_pt_ZZ}]{\includegraphics[width=0.49\linewidth]{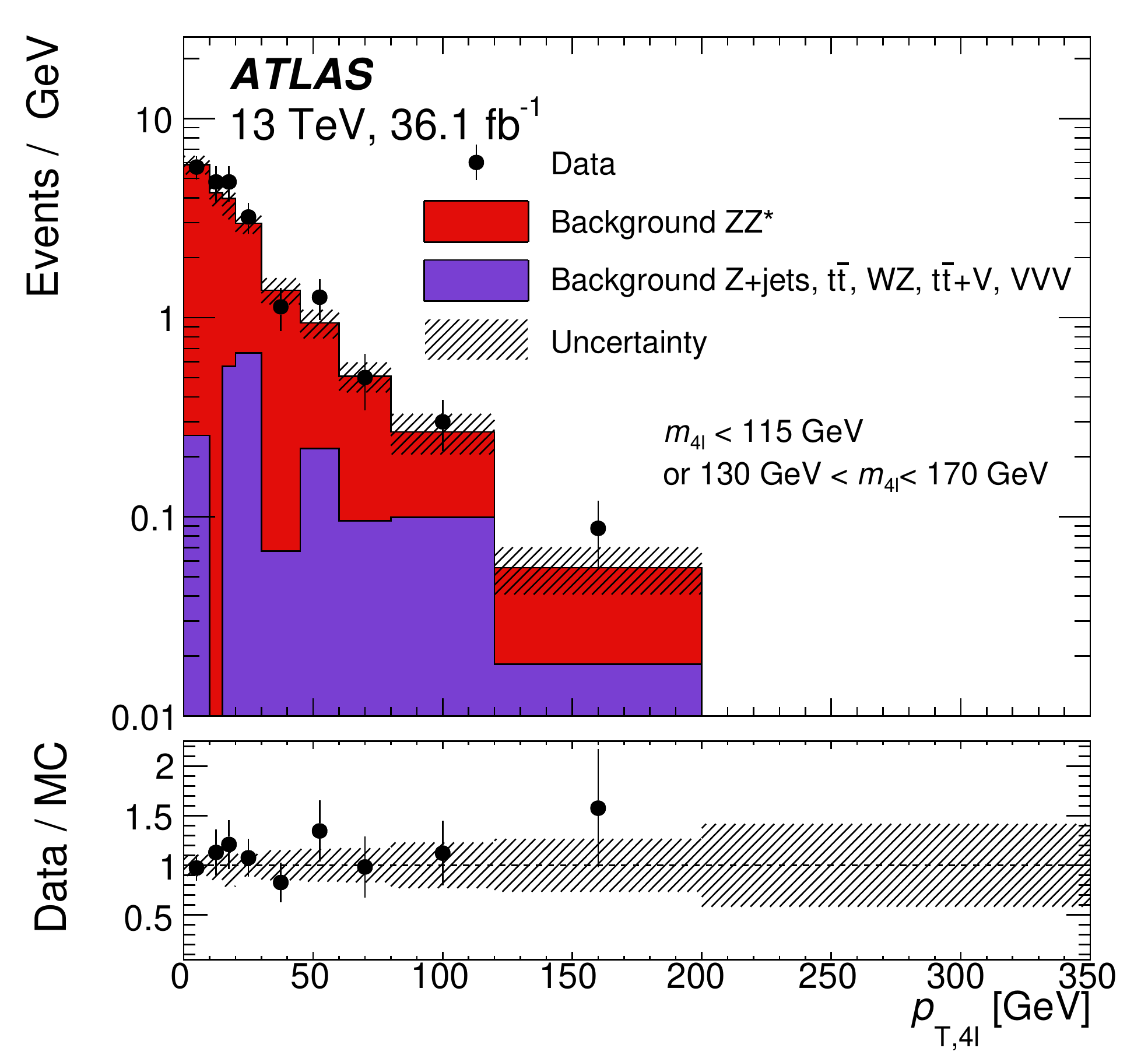}}
\subfigure[\label{fig:cr_njet_ZZ}]{\includegraphics[width=0.49\linewidth]{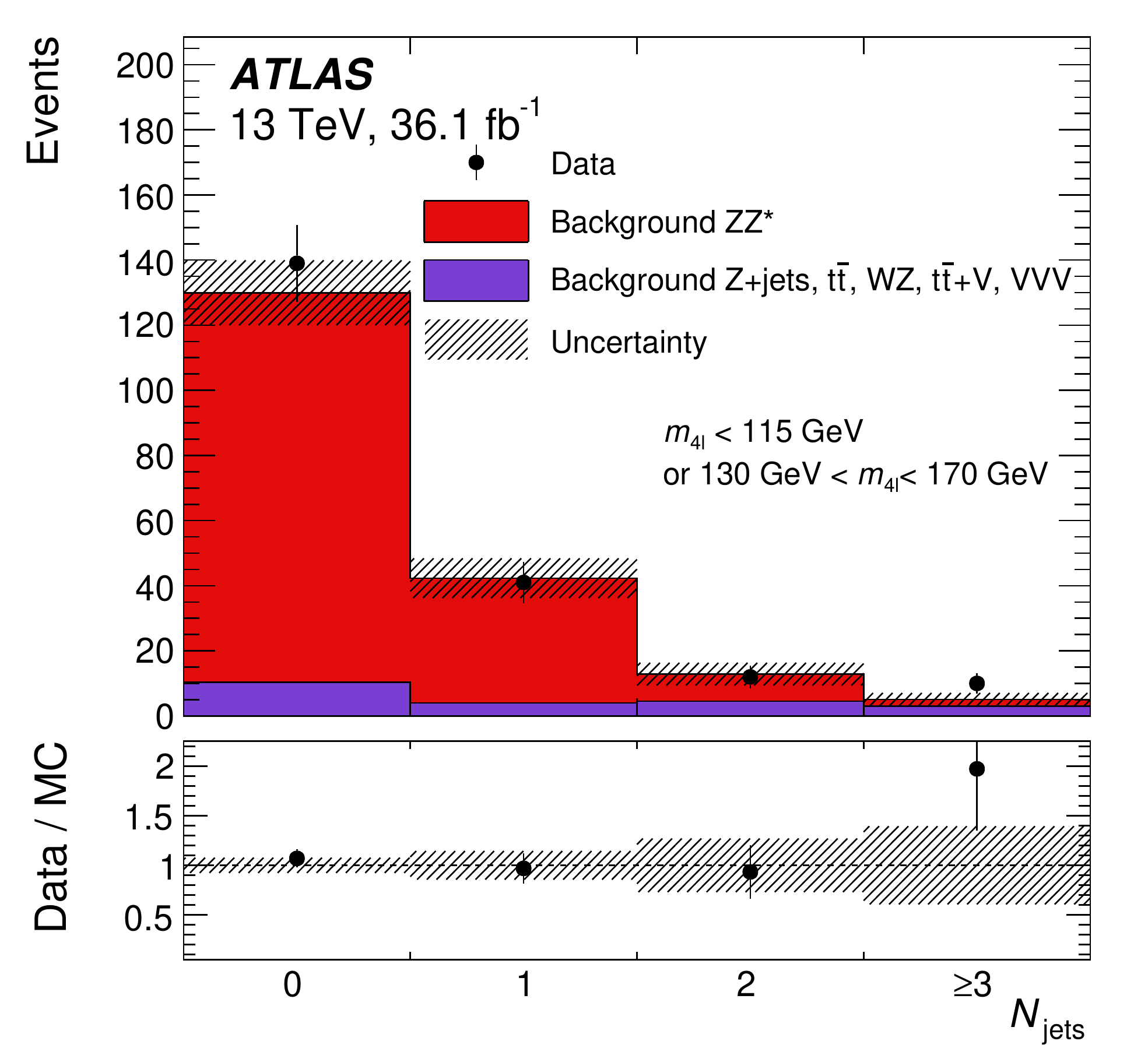}}
\caption{
Reconstructed event yields in bins of (a) the transverse momentum of the four leptons \ptfourl\ and (b) the number of jets \njets, in a non-resonant $ZZ^*$-enriched control region, obtained by applying the full event selection except for the \mfourl\ window, i.e. $m_{4\ell} < 115$~\gev\ or 130~\gev $< m_{4\ell} <$ 170~\gev. The error bars on the data points indicate the statistical uncertainty. The uncertainty in the prediction is shown by the dashed band. The bottom part of the figures shows the ratio of data to the MC expectation.
\label{fig:ZZCR}}
\end{figure}

Other processes that contribute to the background, such as \zjets, \ttbar, and $WZ$, contain at least one jet, photon or lepton candidate that is misidentified as a prompt lepton. These backgrounds are significantly smaller than the non-resonant $ZZ^*$ background and are estimated using data where possible, following slightly different approaches for the \llmumu\ and \llee\ final states~\cite{Aad:2014eva}.

In the \llmumu\ final states, the normalizations for the \zjets\ and \ttbar\ backgrounds are determined using fits to the invariant mass of the leading lepton pair in dedicated data control regions.
The control regions are formed by relaxing the $\chi^2$ requirement on the vertex fit, and by inverting or relaxing isolation and/or impact-parameter requirements on the subleading muon pair. An additional control region ($e\mu \mu\mu$) is used to improve the \ttbar\ background estimate. Transfer factors to extrapolate from the control regions to the signal region are obtained separately for \ttbar\ and \zjets\ using simulation.
The shapes of the \zjets\ and \ttbar\ backgrounds for the differential observables are taken from simulation and normalized using the inclusive data-driven estimate. Comparisons in the control regions show good agreement between data and the simulation for the different observables.

The \llee\ control-region selection requires the electrons in the subleading lepton pair to have the same charge, and relaxes the identification and isolation requirements on the electron candidate with the lowest transverse energy. This electron candidate, denoted as $X$,  can be a light-flavour jet, a photon conversion or an electron from heavy-flavour hadron decay.
The heavy-flavour background is completely determined from simulation, whereas the light-flavour and photon conversion background is obtained with the sPlot~\cite{splot} method, based on a fit to the number of hits in the innermost ID layer  in the data control region. Transfer factors for the light-flavour jets and converted photons, obtained from simulated samples, are corrected using $Z+X$ control regions and then used to extrapolate the extracted yields to the signal region.
Both the extraction of the yield in the control region and the extrapolation are performed in bins of the transverse momentum of the electron candidate and the jet multiplicity.
In order to extract the shape of the backgrounds from light-flavour jets and photon conversions in bins of the differential distributions, a similar method is used, except that the extraction and extrapolation is now performed as a function of the transverse momentum of the electron candidate in each bin of the variable of interest.

The \mfourl\ shapes are extracted from simulation for most background components except for the light-flavour jet + conversion contribution in the \llee\ final state, which is not well described by the simulation and therefore taken from the control region and extrapolated using the data-corrected efficiencies. It was observed that the \mfourl\ shape of the \zjets\ and \ttbar\ backgrounds does not change significantly across the differential distributions, and so the same shape, obtained using all available events, is used for all bins. 

The background from $WZ$ production is included in the data-driven estimates for the \llee\ final
states, while it is added from simulation for the \llmumu\ final states. The contributions from \ttZ\ and triboson processes are very small and taken from simulated samples.

\section{Measured data yields}
\label{sec:yields}

The observed number of events in the four decay channels after the event selection, as well as the expected signal and background yields, is presented in Table~\ref{tab:recoyields}.
Figure~\ref{fig:recoyields} shows the expected and observed event yields for four of the measured differential spectra.
The total observed and predicted event counts agree within 1.3 standard deviations.

\begin{table*}[htbb!]
\centering 	
\caption{Number of expected and observed events in the four decay channels after the event selection, in the mass range \hbox{115~\gev$< \mfourl <$ 130~\gev}. The sum of the expected number of SM Higgs boson events and the estimated background yields is compared to the data. Combined statistical and systematic uncertainties are included for the predictions (see Section~\ref{sec:systematics}). \label{tab:recoyields}}

\vspace{0.2cm}
\begin{tabular}{lcS[table-format=-1.,table-figures-uncertainty=1]ccr}
        \hline \hline
        \noalign{\vspace{0.05cm}}
        Final state  & SM Higgs & \zzstar & \zjets,~$t\bar{t}$ & Expected  & Observed \\
                 &   &         & $WZ$, $ttV$, $VVV$            &                        & \\
\hline
                      $4\mu$  & $ 20.1 \pm   1.6 $ &    9.8\pm   0.8  & $  1.3 \pm  0.3 $ & $  31.2 \pm   1.8 $ & 33 \\
                      $4e$  &  $ 10.6 \pm   1.0 $ &    4.4\pm   0.4  & $  1.3 \pm  0.2 $ & $  16.3 \pm   1.1 $ & 16 \\
                      $2e2\mu$  & $  14.2 \pm   1.1 $  &    7.1\pm   0.5  & $  1.0 \pm  0.2 $ &  $ 22.3 \pm   1.2 $ & 32 \\
                      $2\mu2e$  & $  10.8 \pm   1.0 $ &   4.6\pm   0.5  & $  1.4 \pm  0.3 $ & $  16.8 \pm   1.1 $  & 21 \\
                      \noalign{\vspace{0.05cm}}
        \hline
                         Total  &   $  56 \pm     4 $  &   25.9\pm   2.0  & $   5.0 \pm   0.7 $ & $    87 \pm     5 $ & 102 \\
       \hline \hline 
    \end{tabular}
\end{table*}

\begin{figure}[!htbp]
\centering
\subfigure[\label{fig:r_jetpt}]{\includegraphics[width=0.49\linewidth]{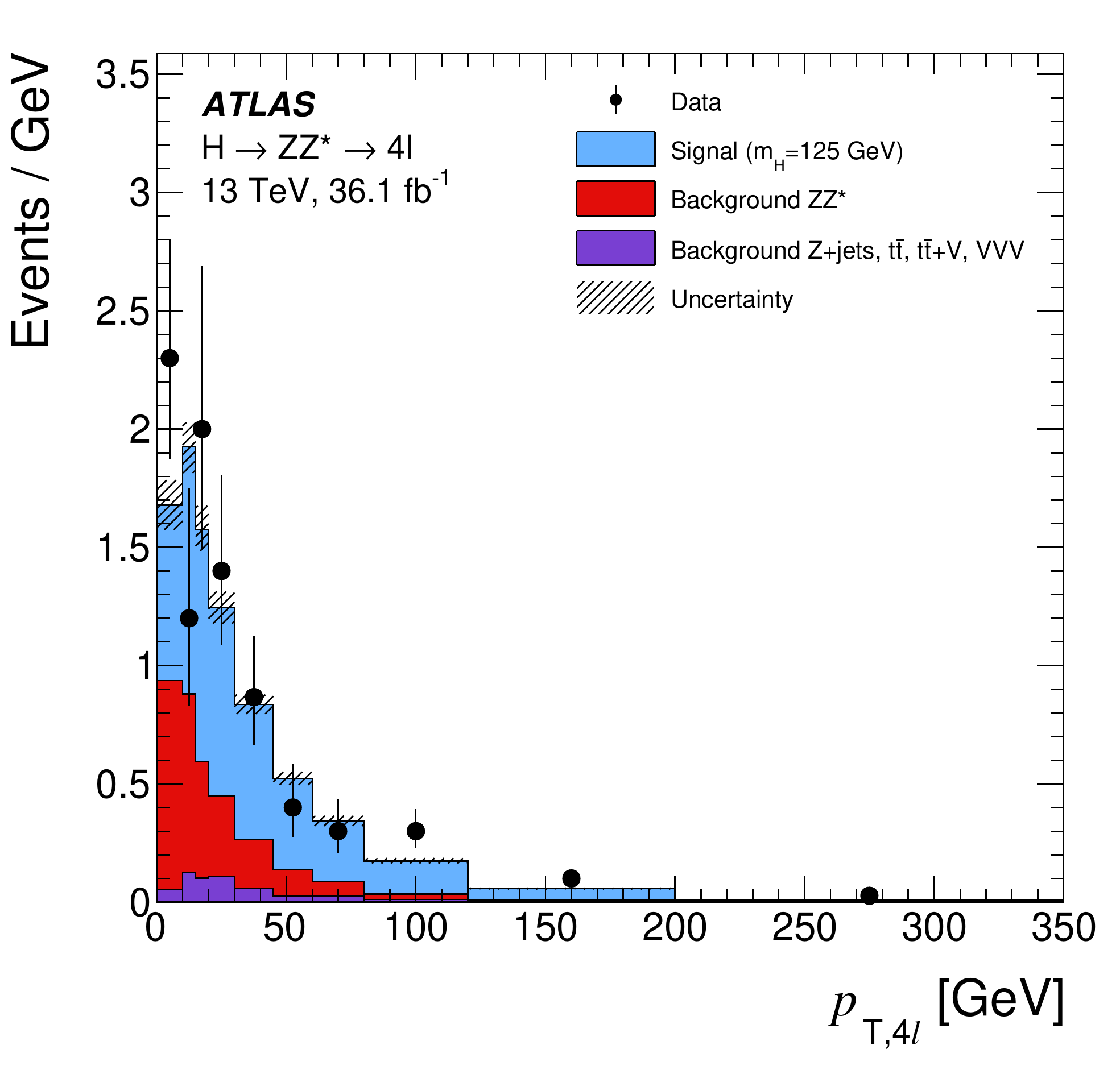}}
\subfigure[\label{fig:r_mjj}]{\includegraphics[width=0.49\linewidth]{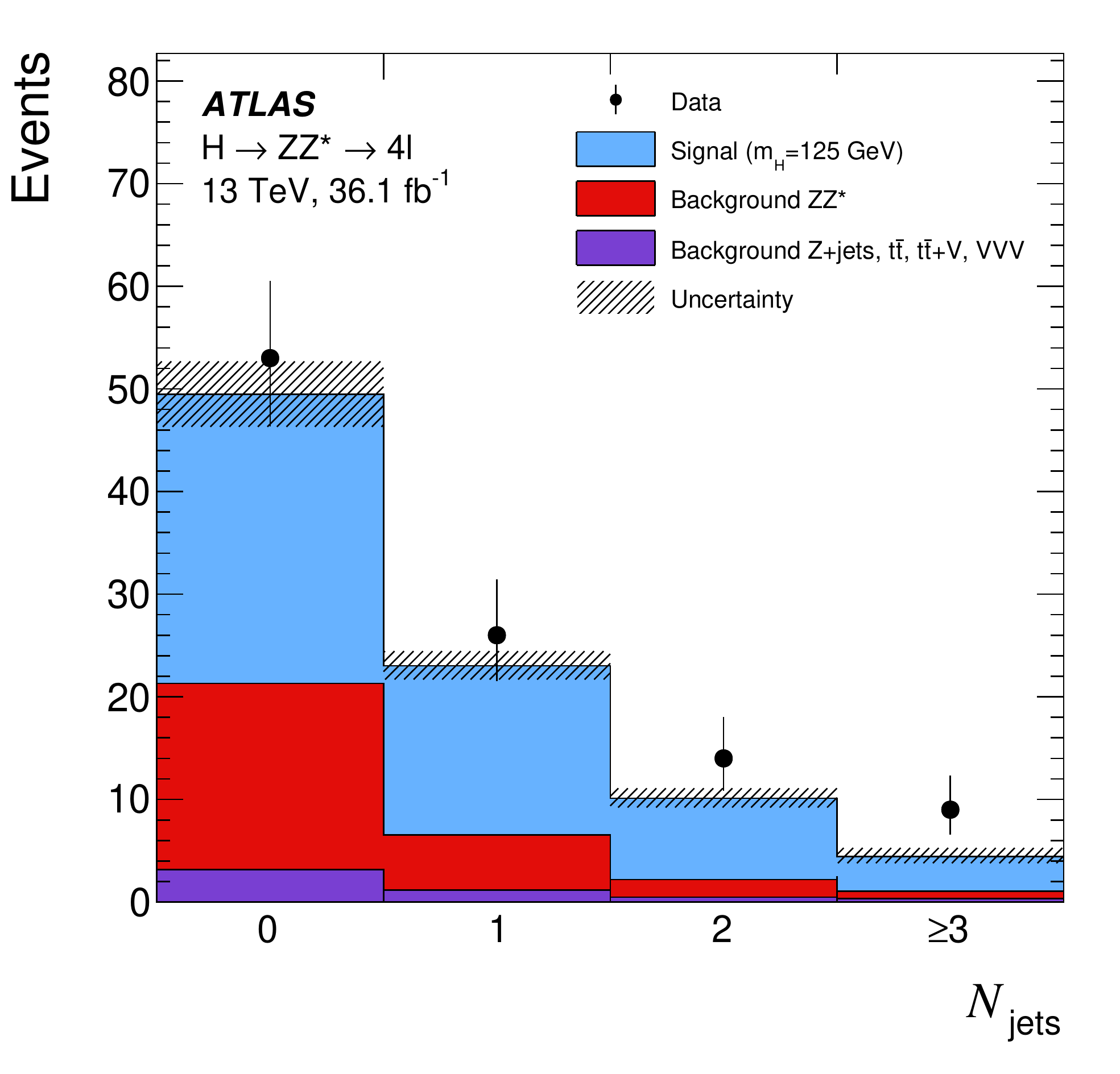}}
\subfigure[\label{fig:r_mjj}]{\includegraphics[width=0.49\linewidth]{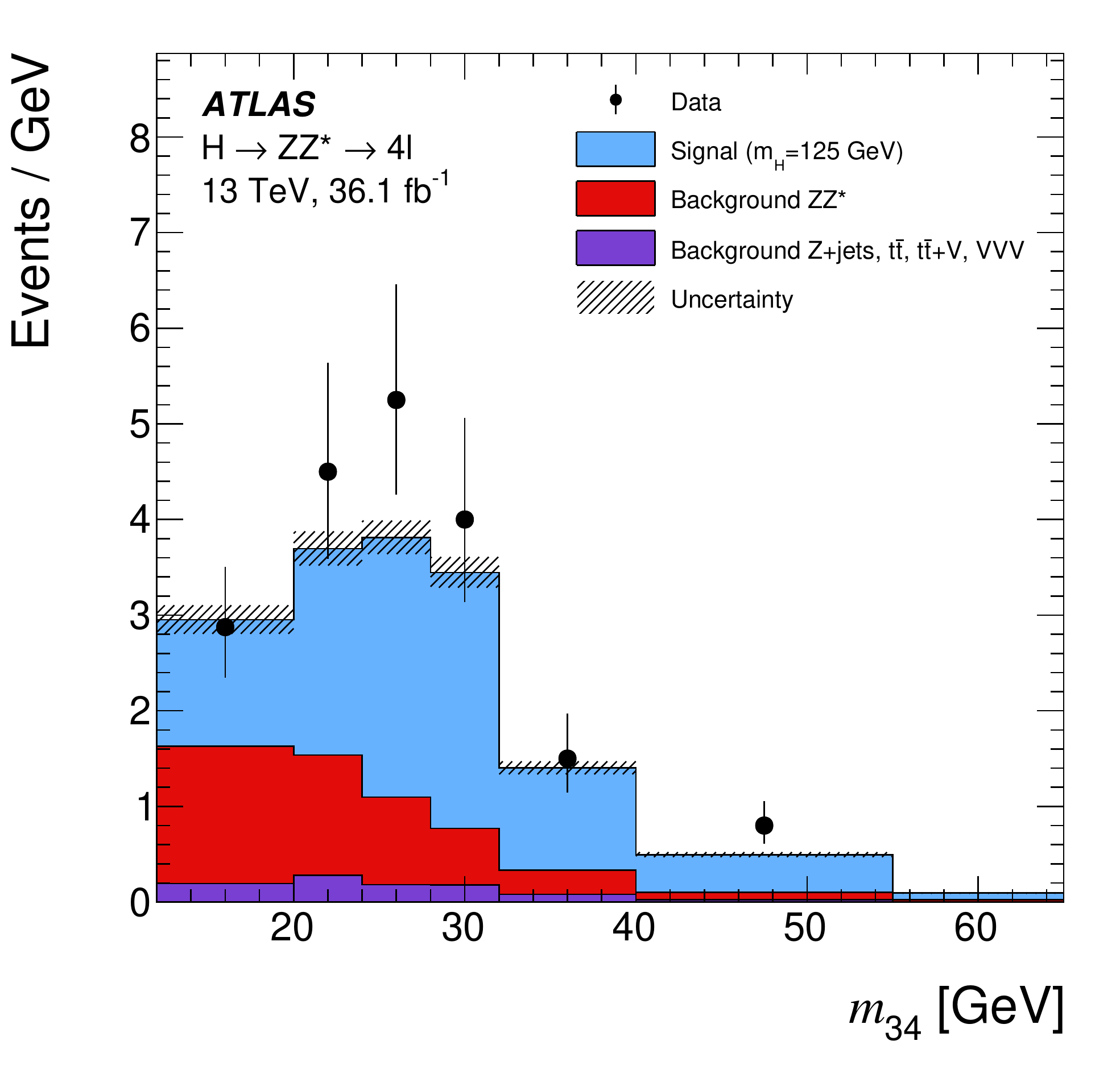}}
\subfigure[\label{fig:r_m12m34}]{\includegraphics[width=0.49\linewidth]{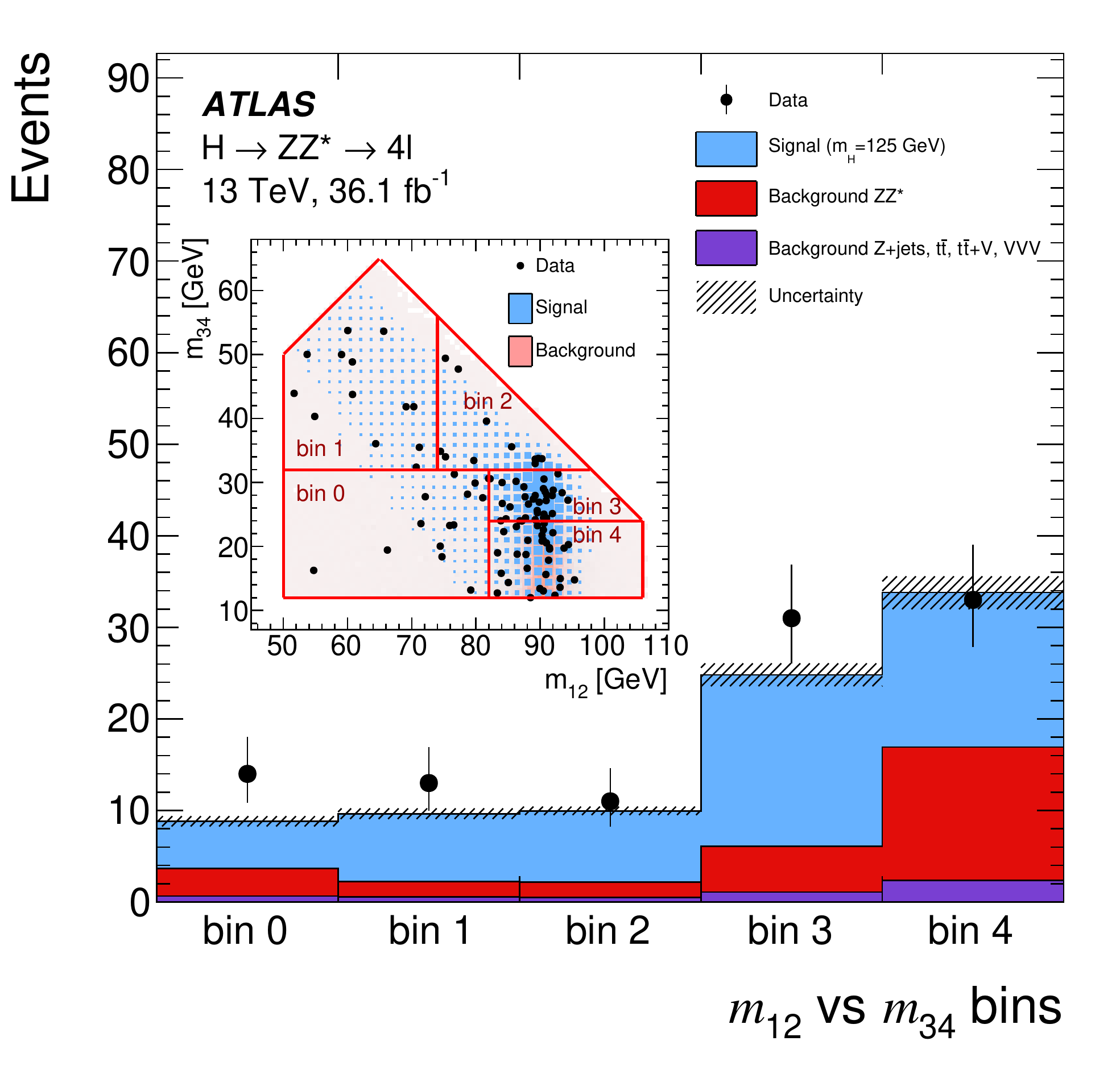}}
\caption{Measured data yields compared to SM Higgs boson signal and background processes for (a) the transverse momentum of the four leptons \ptfourl, (b) the number of jets \njets, (c) the invariant mass of the subleading lepton pair \mthreefour, and (d) the invariant mass of the leading vs the subleading pair \monetwo\ vs \mthreefour. Figure~(d)  also includes an illustration of the chosen bins, as well as the two-dimensional distributions of data and prediction. The error bars on the data points indicate the statistical uncertainty. The  uncertainty in the prediction is shown by the dashed band.  \label{fig:recoyields}}
\end{figure}

\section{Signal extraction and correction for detector effects}
\label{sec:unfolding}

\begin{figure}[!htbp]
\centering
\subfigure[\label{fig:m4l_4mu}]{\includegraphics[width=0.49\linewidth]{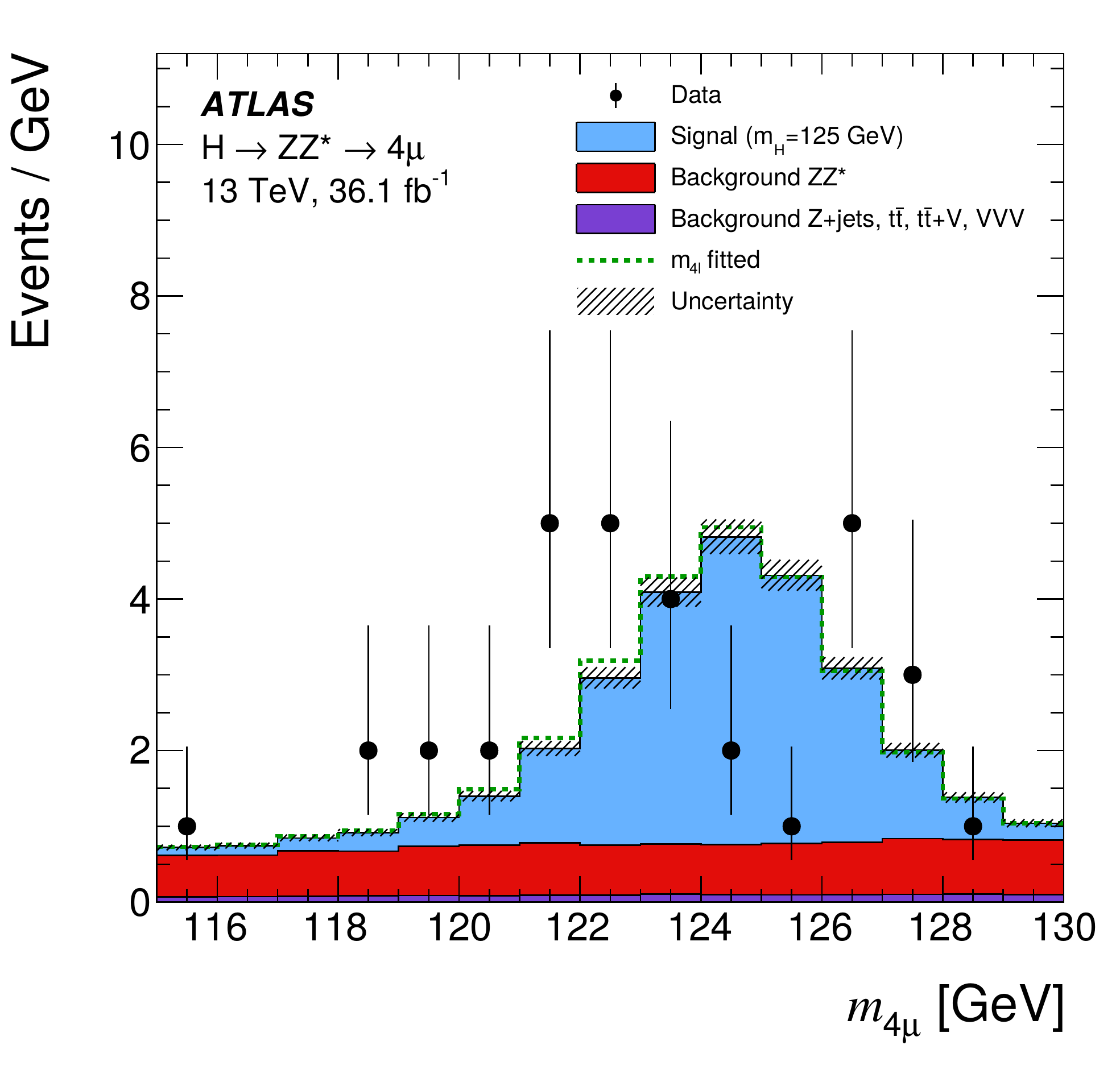}}
\subfigure[\label{fig:m4l_4e}]{\includegraphics[width=0.49\linewidth]{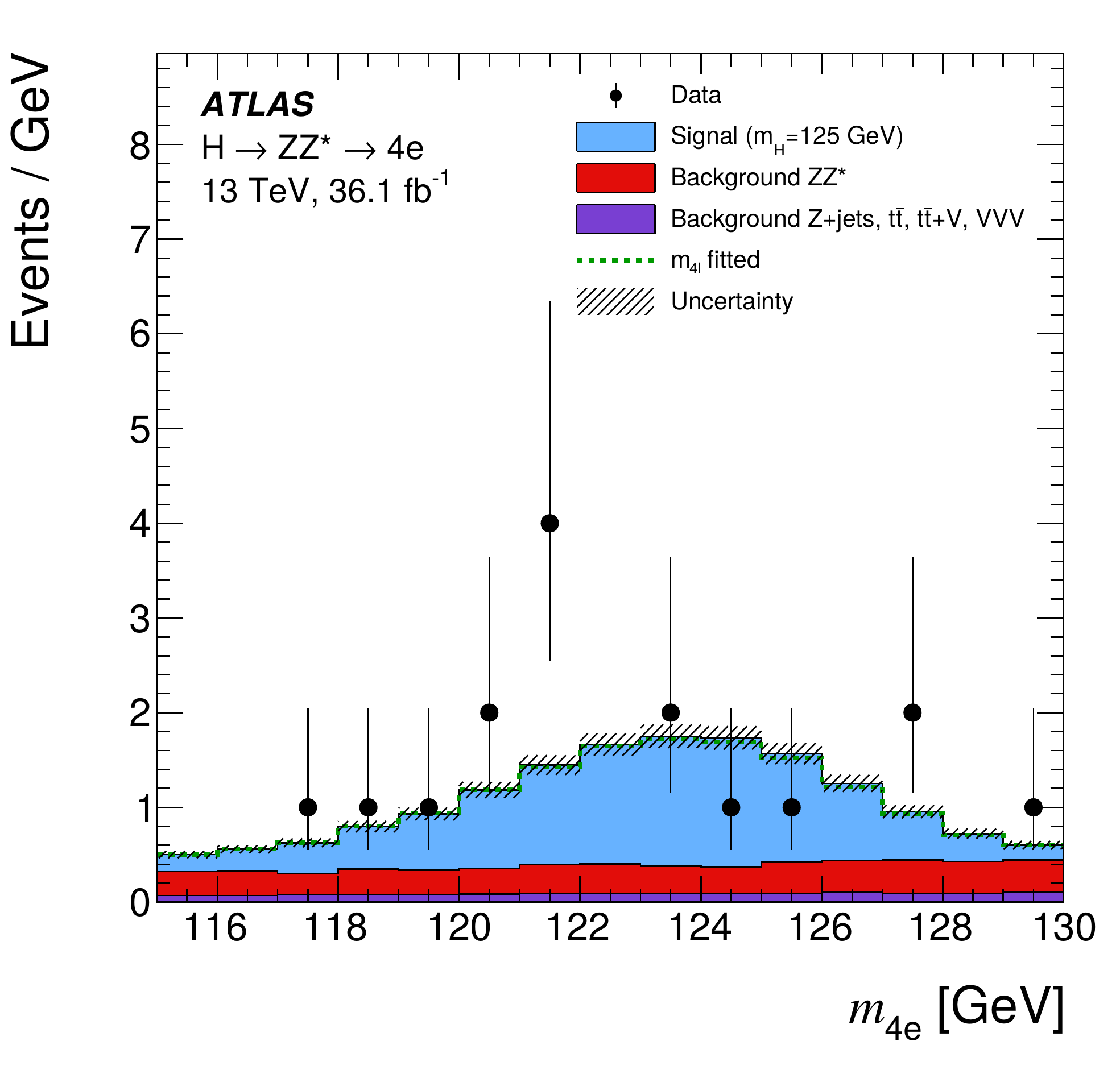}}
\subfigure[\label{fig:m4l_2mu2e}]{\includegraphics[width=0.49\linewidth]{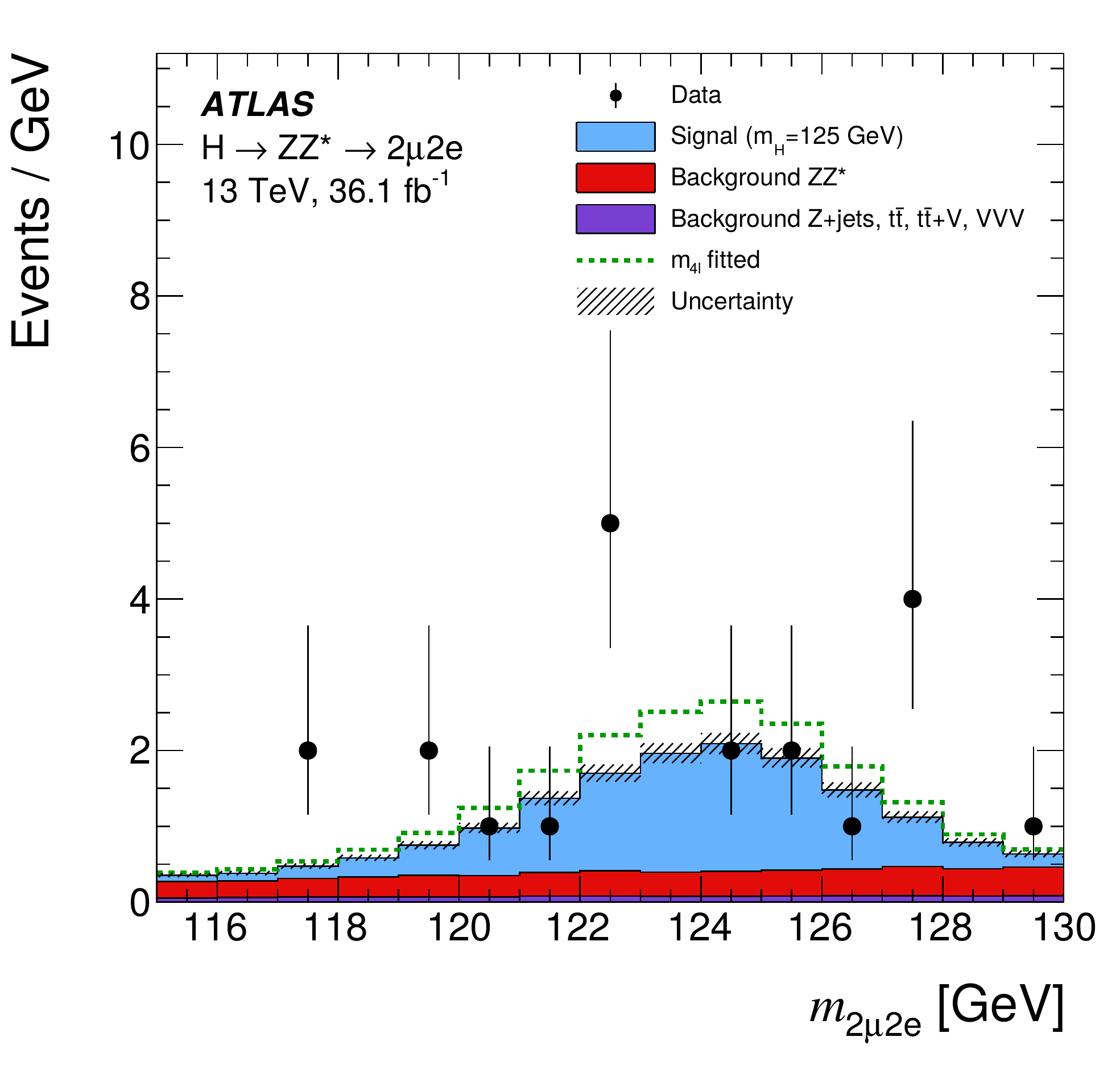}}
\subfigure[\label{fig:m4l_2e2mu}]{\includegraphics[width=0.49\linewidth]{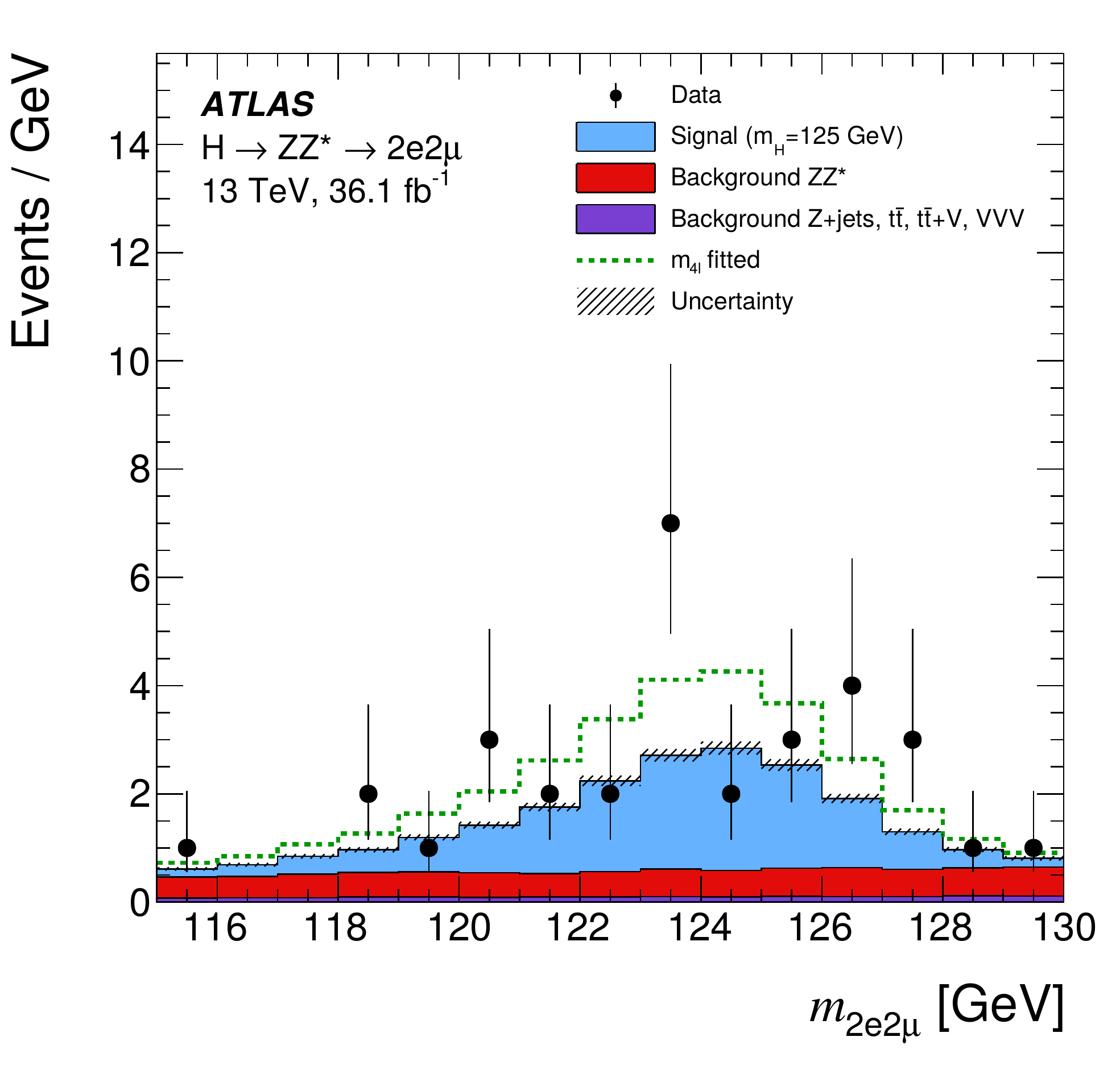}}
\caption{Template fit of SM Higgs boson signal and background to the data for the inclusive distributions for the different decay channels (a) 4$\mu$, (b) 4$e$, (c) $2\mu 2e$, (d) $2e2\mu$. The error bars on the data points indicate the statistical uncertainty. The SM Higgs boson predictions are normalized to the cross sections discussed in Section~\ref{sec:samples}, while the backgrounds are normalized to the estimates described in Section~\ref{sec:signalextraction}. The  uncertainty in the prediction is shown by the dashed band. The dotted green line illustrates the best fit.
\label{fig:mfourellfits_fid} }
\end{figure}

\begin{figure}[!htbp]
\centering
\subfigure[\label{fig:m4l_pt0}]{\includegraphics[width=0.49\linewidth]{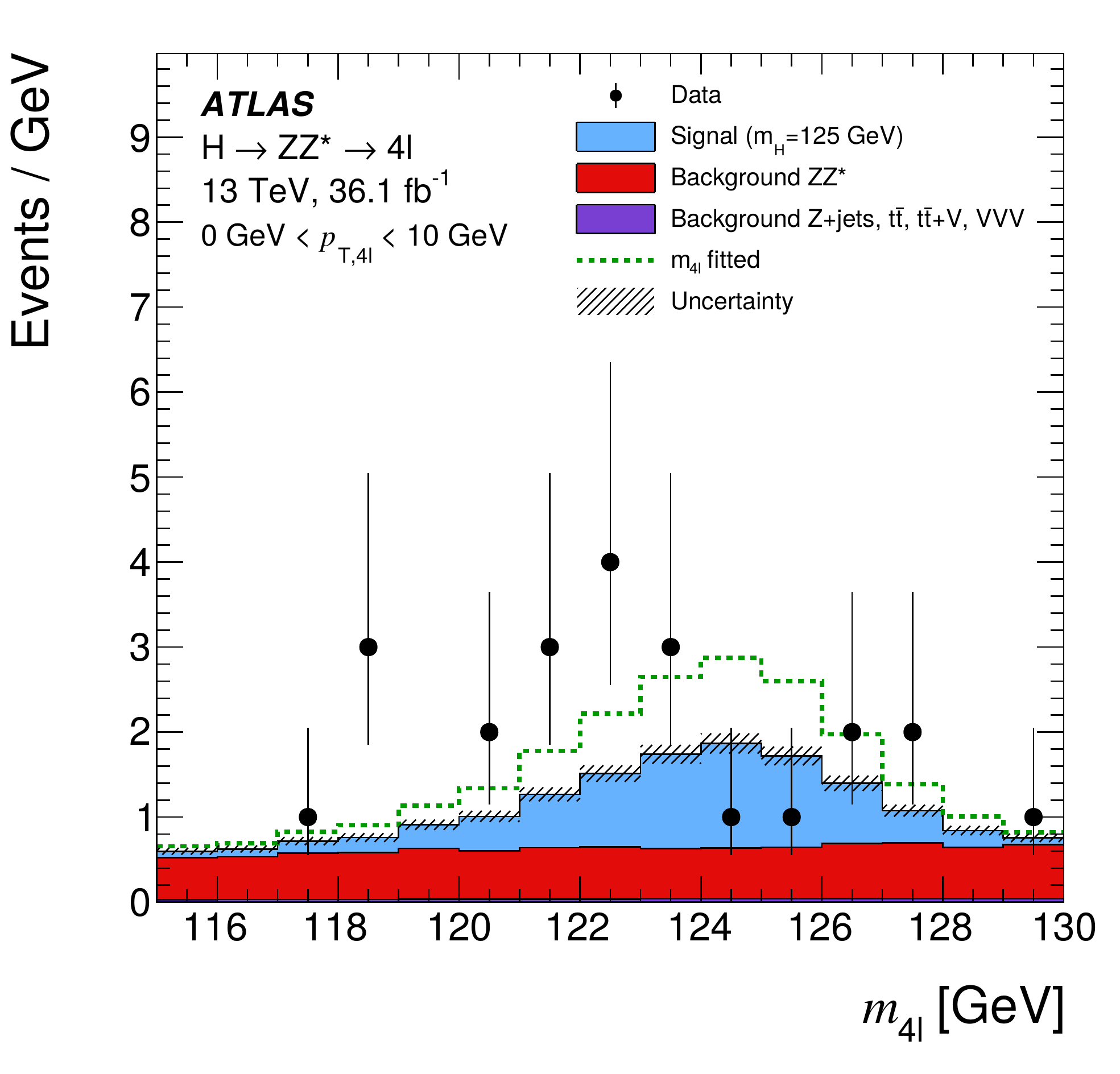}}
\subfigure[\label{fig:m4l_pt9}]{\includegraphics[width=0.49\linewidth]{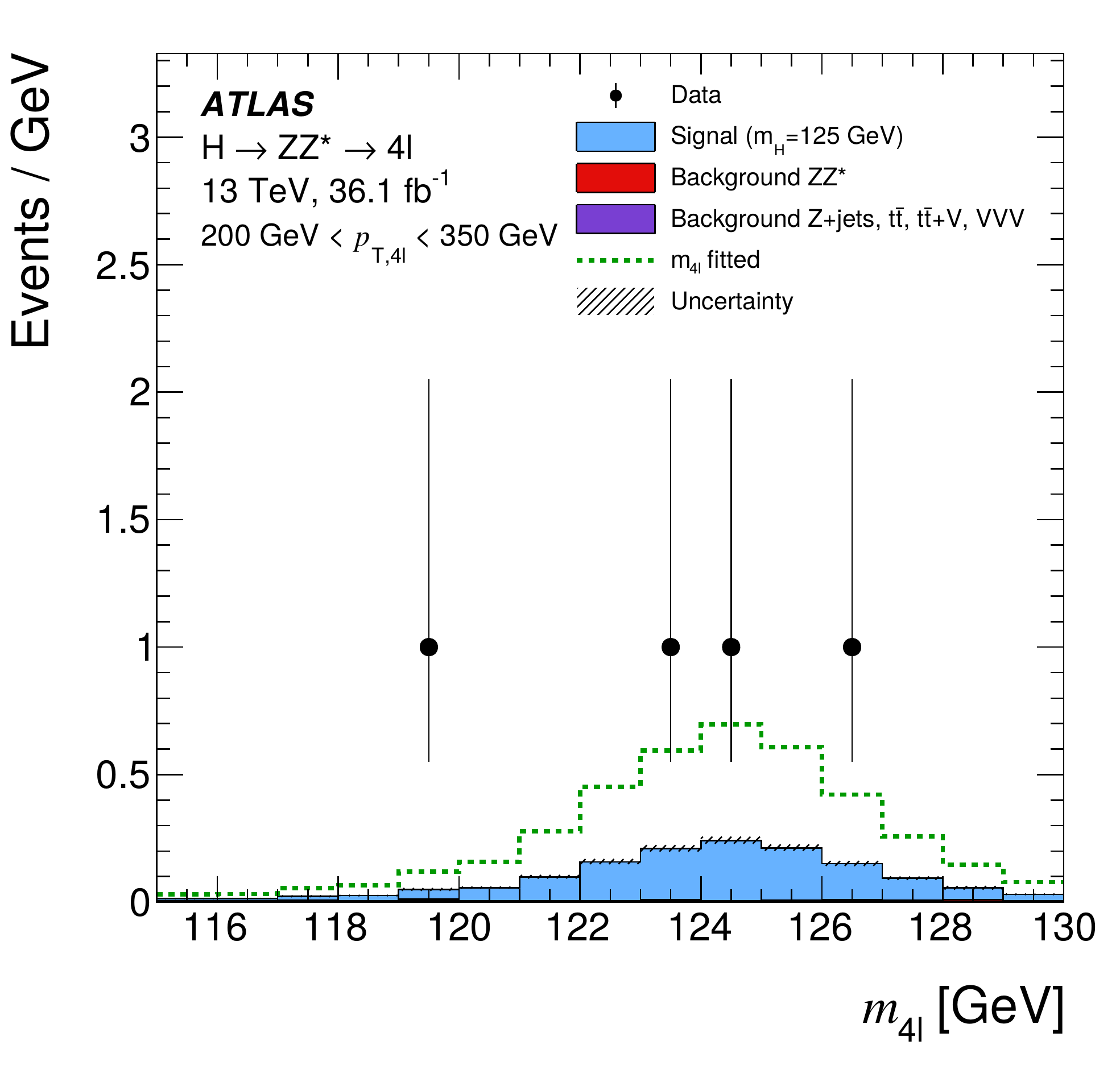}}
\caption{Template fit of SM Higgs boson signal and background to the data for the (a) first and (b) last bins of the distribution of the transverse momentum of the four leptons \ptfourl. The error bars on the data points indicate the statistical uncertainty. The SM Higgs boson predictions are normalized to the cross sections discussed in Section~\ref{sec:samples}, while the backgrounds are normalized to the estimates described in Section~\ref{sec:signalextraction}. The  uncertainty in the prediction is shown by the dashed band. The dotted green line illustrates the best fit.  \label{fig:mfourellfits_pt}}
\end{figure}

To extract the number of signal events in each bin of a differential distribution (or for each decay channel for the inclusive fiducial cross section), invariant mass templates for the Higgs boson signal and the background processes are fit to the \mfourl\ distribution in data. The signal shape is obtained from the simulated samples described in Section~\ref{sec:samples} assuming a Higgs boson mass of 125~\gev. Most of the background shapes are also obtained from the simulated samples described in Section~\ref{sec:samples}, while some of the backgrounds in the \llee\ channel are derived from control regions in data, as discussed in Section~\ref{sec:signalextraction}. The normalization of the backgrounds is fixed in this fit. Figures~\ref{fig:mfourellfits_fid} and \ref{fig:mfourellfits_pt} show the data, templates and best fits for the \mfourl\ distributions in the four decay channels for the extraction of the inclusive fiducial cross section, and two bins of the transverse momentum of the four leptons. For the differential distributions, no split into decay channels is performed, and the SM $\zzstar \rightarrow 4\ell$ decay fractions are assumed.

The fiducial cross section $\sigma_{i,\mathrm{fid}}$ for a given final state or bin of the differential distribution is defined as:

\begin{align}
\label{eq:fidXS}
 \sigma_{i,\mathrm{fid}}= \sigma_i \times A_i \times \mathcal{B} = \frac{N_{i,\mathrm{fit}}}{\mathcal{L} \times C_i }, ~~~ C_i = \frac{N_{i,\mathrm{reco}}}{N_{i,\mathrm{part}}},
\end{align}

where $A_i$ is the acceptance in the fiducial phase space, $\mathcal{B}$ is the branching ratio and $\sigma_i$ is the total cross section in bin $i$. The term $N_{i,\mathrm{fit}}$ is the number of extracted signal events in data, $\mathcal{L}$ is the integrated luminosity and $C_i$ is the bin-by-bin correction factor for detector inefficiency and resolution. The term $N_{i,\mathrm{reco}}$ is the number of reconstructed signal events and $N_{i,\mathrm{part}}$ is the number of events at the particle level in the fiducial phase-space.
The correction factor is calculated from simulated Higgs boson samples, assuming SM production mode fractions and $\zzstar \rightarrow 4\ell$ decay fractions as discussed in Section~\ref{sec:samples}. The systematic uncertainties in this assumption are described in Section~\ref{sec:systematics}.
The correction factors for the different Higgs boson production modes agree within 15\%, except for the \ttH\ mode, which differs by up to 40\%, due to the fact that \ttH\ events have more hadronic jets and that no isolation requirements are applied to the leptons at the particle level. 
The correction factors for the four final states are 0.64 $\pm$ 0.04 (4$\mu$), 0.55 $\pm$ 0.03 ($2e2\mu$), 0.48 $\pm$ 0.05 (2$\mu 2e$), and 0.43 $\pm$ 0.06 (4$e$).
Figure~\ref{fig:corrfactors} shows the bin-by-bin correction factors for all decay channels combined including systematic uncertainties for the \ptfourl\ and \njets\ distributions. The large uncertainty for  \njets $\ge$ 3 is due to the experimental jet reconstruction uncertainties and the variations of the fractions of Higgs boson production modes (see Section~\ref{sec:systematics}).
The same figure also shows the bin purity, defined as the fraction of events in a bin of the reconstructed distribution that are found in the same bin at particle level. The bin purity is greater than $0.75$ for the Higgs boson kinematic and decay observables, and typically greater than $0.6$ for the jet variables.
It can be seen that the narrower bins at low \ptfourl\ have a slightly reduced bin purity, as detector resolution effects result in larger bin migration effects, which is enhanced by the presence of a steep slope.

\begin{figure}[!htbp]
\centering
\subfigure[\label{fig:c_pt}]{\includegraphics[width=0.49\linewidth]{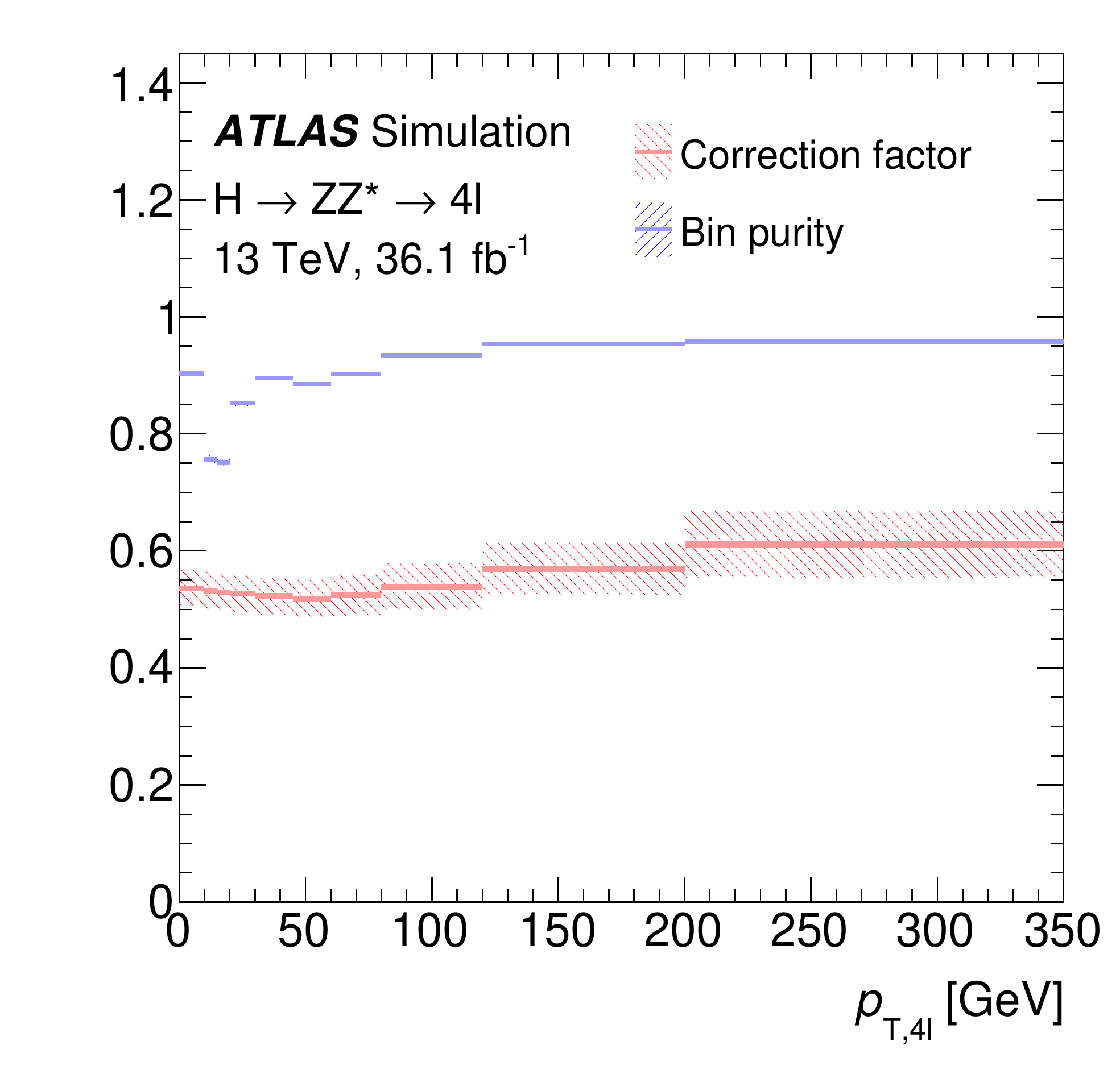}}
\subfigure[\label{fig:c_njets}]{\includegraphics[width=0.49\linewidth]{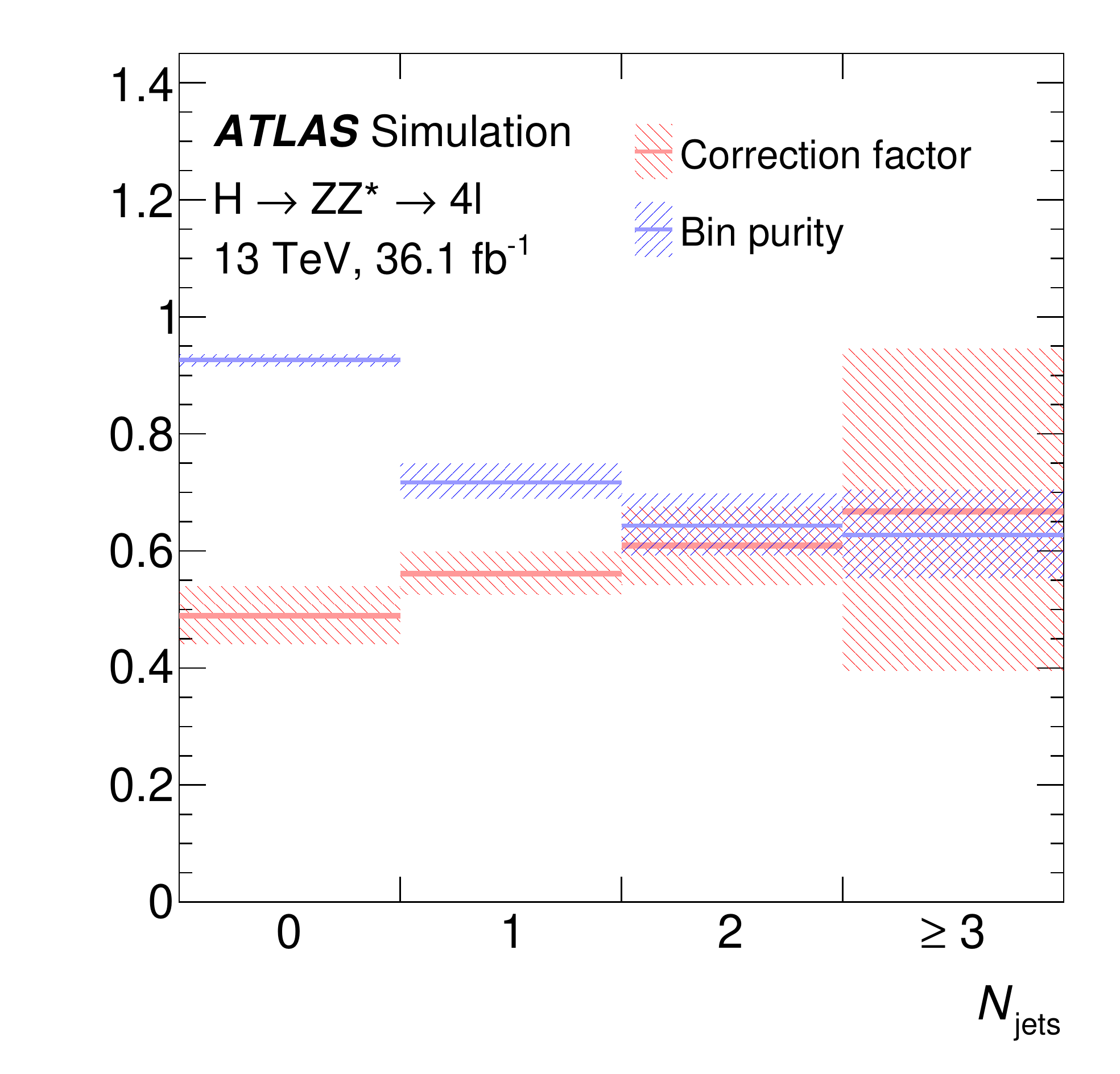}}
\caption{Bin-by-bin correction factors and bin purities for (a) the transverse momentum of the four leptons \ptfourl\ and (b) the number of jets \njets. The bands show the systematic uncertainties in the correction factors, which are discussed in Section~\ref{sec:systematics}. The uncertainties in the bin purity include the detector response and pile-up uncertainties. \label{fig:corrfactors}}
\end{figure}

The signal, background, and data \mfourl\ distributions, as well as the correction factors, are used as input to a profile-likelihood-ratio fit~\cite{Cowan:2010js}, taking into account all bins of a given distribution and all final states for the inclusive measurement. The likelihood includes the shape and normalization uncertainties of the backgrounds and correction factors as nuisance parameters.  This allows for correlation of systematic uncertainties between the background estimates and the correction factors, as well as between bins or decay channels. The cross sections are extracted for each bin, or final state, by minimizing twice the negative logarithm of the profile likelihood ratio, $-2 \ln{\Lambda}$. In the asymptotic assumption, i.e. the large sample limit, $-2 \ln{\Lambda}$ behaves as a $\chi^2$ distribution with one degree of freedom. The compatibility of a measured cross section and a theoretical prediction is evaluated by computing a $p$-value based on the difference between the value of $-2 \ln{\Lambda}$ at the best-fit value and the value obtained by fixing the cross sections in all bins to the ones predicted by the theory.
These $p$-values do not include the uncertainties in the theoretical predictions, which are significantly smaller than the total data uncertainties. Therefore, they are slightly smaller than they would be with all uncertainties included. For all measured observables the asymptotic assumption is verified with pseudo-experiments, and if necessary, the uncertainties are corrected to the values obtained with the pseudo-experiments. In the case of zero observed events, 95\% confidence level (CL) limits on the fiducial cross sections are set using the CL$_\mathrm{s}$ modified frequentist formalism~\cite{cls,Cowan:2010js}.

The inclusive fiducial cross section for each channel is calculated from the fit results following Eq.~(\ref{eq:fidXS}). The fiducial cross sections of the four final states can either be summed together to obtain an inclusive fiducial cross section, or they can be combined assuming the SM $\zzstar \rightarrow 4\ell$ branching ratios. The latter combination is more model dependent, but benefits from a smaller statistical uncertainty.

\section{Systematic uncertainties}
\label{sec:systematics}

Experimental systematic uncertainties affecting both the simulated background and correction factors arise from uncertainties in the efficiencies, resolutions and energy scales of leptons and jets~\cite{Aad:2016jkr,ATLAS-CONF-2016-024,ref:run1-egamma-calib,Aaboud:2017jcu}, as well as pile-up modelling. These uncertainties can affect both the shape and the normalization of the distributions.
For the background estimate and the conversion of the corrected signal yields to cross sections, the luminosity uncertainty needs to be taken into account. The uncertainty in the combined 2015+2016 integrated luminosity is 3.2\%, which affects the signal and simulated background estimates. It is derived, following a methodology similar to that detailed in Ref.~\cite{Aaboud:2016hhf}, from a preliminary calibration of the luminosity scale using x--y beam-separation scans performed in August 2015 and May 2016.

Uncertainties  in the estimation of \zjets, \ttbar, and $WZ$ backgrounds are also considered.  The dominant systematic uncertainties here arise from difficulties in modelling the extrapolation from the control regions to the signal region, which can affect not only the overall normalization but also the background composition estimates and hence the yields in the bins of the differential distributions.

For the simulated backgrounds and the extrapolation of the inclusive fiducial cross section to the total cross section, theoretical modelling uncertainties associated with PDF, missing higher-order QCD corrections (via variations of the factorization and renormalization scales), as well as underlying event and parton showering uncertainties are considered. For the extrapolation to the total cross section, uncertainties in the \hfourl\ branching ratios are also included~\cite{Dittmaier:2011ti}.

The effect on the fitted event yields of shifting the \mfourl\ template according to the uncertainties in the measured Higgs boson mass, 0.24~\gev~\cite{mass}, is smaller than 0.5\% and therefore neglected.

The dependence of the correction for detector effects on the theoretical modelling is assessed in a number of ways. For ggF, VBF and $VH$, the PDF4LHC NLO PDF set is varied according to its eigenvectors, and the envelope of the variations is used as the systematic uncertainty. The renormalization and factorization scales are varied by factors of 2.0 and 0.5.
Furthermore, \mH\ is varied within the uncertainties in the measured Higgs mass.
The relative contribution of each Higgs boson production mechanism is varied by an amount consistent with the uncertainties obtained from the combined ATLAS and CMS measurement of the Higgs boson production cross sections~\cite{CombCouplings}, except for \ttH\ where the allowed variation is inflated to cover the measured value, which is more than two standard deviations away from the SM prediction. The correction factors are cross-checked using the alternative \progname{Madgraph5} ggF samples (for SM and modified couplings) and the differences with respect to nominal values are found to be well within the statistical uncertainties of the samples. Bias studies and cross-checks with other unfolding methods, such as matrix inversion and Bayesian iterative unfolding~\cite{DAgostini:1994fjx} show results that agree very well with the bin-by-bin correction factor results. Observed differences are generally much smaller than the statistical uncertainties. 

The uncertainties in this analysis are dominated by the limited number of data events. The statistical uncertainty in the fiducial inclusive cross section obtained by combining all decay channels is 14\%, while the systematic uncertainty is 7\%, dominated by the lepton uncertainties and the uncertainty in the luminosity.
For the differential cross sections, the size of the statistical and systematic uncertainties depends on the variable and is shown in Table~\ref{tab:SysUnc}. The breakdown of the dominant systematic uncertainties is obtained by performing the fits while fixing groups of nuisance parameters to their best-fit value. The statistical uncertainties are mostly in the range 20$-$50\%, and can be as high as 150\%.
For the Higgs boson kinematic properties, the most important systematic uncertainties are the experimental lepton uncertainties, 1$-$5\%.
The signal composition uncertainty grows with the increase of the fraction of \ttH\ in some regions of phase space. Therefore, for observables defined by the jet activity produced in association with the Higgs boson, not only the jet energy scale but also the signal composition uncertainties become increasingly important, especially at high \njets\ and \jetpt\ ($\sim$20\% each for \njets $\geq$ 3).

\begin{table*}
\centering
\caption{Fractional uncertainties for the inclusive fiducial cross section $\sigma_\mathrm{comb}$, obtained by combining all decay channels,  and ranges of systematic uncertainties for the differential observables. The columns $e$, $\mu$, jets represent the experimental uncertainties in lepton and jet reconstruction and identification. The $ZZ^*$ theory uncertainties include the PDF and scale variations. The model uncertainties are dominated by the production mode composition variations in the extraction of the correction factors.
\label{tab:SysUnc} }
\vspace{0.2cm}
\small
\begin{tabular}{ ld{-1}d{-1}d{-1}d{-1}d{-1}d{-1}d{-1}d{-1}d{-1} }
\hline
Observable  & \multicolumn{1}{c}{Stat}   & \multicolumn{1}{c}{Systematic}  & \multicolumn{7}{c}{Dominant systematic components [\%] }  \\
     &  \multicolumn{1}{c}{unc. [\%]} & \multicolumn{1}{c}{unc. [\%]} & \multicolumn{1}{c}{$e$} & \multicolumn{1}{c}{$\mu$} & \multicolumn{1}{c}{jets} & \multicolumn{1}{c}{$ZZ^*$ theo} & \multicolumn{1}{c}{Model} & \multicolumn{1}{c}{\zjets + $t\bar{t}$} & \multicolumn{1}{c}{Lumi}\\
\hline
$\sigma_\mathrm{comb}$                           &           \multicolumn{1}{c}{14}   &  \multicolumn{1}{c}{7}  &  \multicolumn{1}{c}{3}  &  \multicolumn{1}{c}{3}   &   \multicolumn{1}{c}{$<0.5$} &  \multicolumn{1}{c}{2}  &  \multicolumn{1}{c}{0.8}  & \multicolumn{1}{c}{0.8} &  \multicolumn{1}{c}{4} \\
d$\sigma$ / d\ptH                          &  30-150   &  3-11  &  1-4  &  1-3    & \multicolumn{1}{c}{$<0.5$}   &  \multicolumn{1}{c}{$<7$}  &  \multicolumn{1}{c}{$<6$}  & 1-6 & 3-5 \\
d$\sigma$ / d\ptH\ (0j)    &  31-52   &  10-18  &  2-5  &  1-4  & 3-16  &  3-8  &  1  & 2-3 & 3-5 \\
d$\sigma$ / d\ptH\ (1j)    &  35-15  &  6-30   &  1-4  &  1-3   & 2-29  &  1-4  &  1-11 & 1-2 & 3-5 \\
d$\sigma$ / d\ptH\ (2j)   &  30-41   &  5-21  &  1-3  &  1-3  &  2-19  &  1-5  &  1-7  & 1-2 & 3-5 \\
d$\sigma$ / d\yH                           &  29-120   &  5-8    &  2-4  &  2-3   & \multicolumn{1}{c}{$<0.5$}    &  1-2  &  \multicolumn{1}{c}{$<1$}  & \multicolumn{1}{c}{1} & 3-5 \\
d$\sigma$ / d\costhetastar                 &  31-100   &  5-8   &  2-4  &  2-3    &  \multicolumn{1}{c}{$<0.5$}  &  1-2  &  \multicolumn{1}{c}{$<2$}  & 1-4 & 3-5 \\
d$\sigma$ / d\mthreefour                    &  26-53   &  4-13  &  2-5  &  1-5    & \multicolumn{1}{c}{$<0.5$}   &  1-6  &  \multicolumn{1}{c}{$<1$}  & 1-3 & 3-5 \\
d$^2\sigma$ / d\monetwo d\mthreefour    &  21-40   &  4-12  &  2-4  &  1-4    &  \multicolumn{1}{c}{$<0.5$}  &  1-6  &  \multicolumn{1}{c}{$<1$}  & 1-4 & 3-5 \\
d$\sigma$ / d\njets                         &  22-44   &  6-31  &  1-4  &  1-3   & 4-22  &  2-4  &  1-22 & 1-2 & 3-5 \\
d$\sigma$ / d\jetpt                         &  30-53   &  5-18  &  1-4  &  1-3  &  3-16  &  2-3  &  1-8  & 1-2 & 3-5 \\
d$\sigma$ / d\dphijj                        &  29-43   &  9-17  &  1-3  &  1-3  &  8-14 &  3-4  &  1-7  & 1 & 3-5 \\
d$\sigma$ / d\mjj                           &  23-100  &  9-27  &  1-4  &  1-4   & 8-24  &  3-8  &  1-7  & \multicolumn{1}{c}{$<3$} & 3-5 \\
\hline
\hline
\end{tabular}
\end{table*}

\section{Results}
\label{sec:results}

The inclusive fiducial cross sections of \hfourl\ are presented in Table~\ref{tab:FiducialResults} and Figure~\ref{fig:unfolded_fid}. The left panel in Figure~\ref{fig:unfolded_fid} shows the fiducial cross sections for the four individual decay channels ($4\mu$, $4e$, $2\mu 2e$, $2e2\mu$). The middle panel shows the cross sections for opposite- and same-flavour decays, which can provide a handle on same-flavour interference effects, as well as the fiducial cross sections obtained by either summing all 4$\ell$ decay channels or combining them assuming SM branching ratios.
The data are compared to the LHCXSWG prediction after accounting for the fiducial acceptance as determined from the SM Higgs boson simulated samples (see Section~\ref{sec:samples}).
The fiducial cross section is extrapolated to the total phase space, as shown in the right panel, using the same fiducial acceptance as well as the branching ratios, with the additional uncertainties described in Section~\ref{sec:systematics}. The total cross section is also compared to the cross sections predicted by \progname{NNLOPS}, \progname{HRes}, and \mgfxfx~(see Section \ref{sec:samples}). It can be seen that the \mgfxfx\ cross section is lower than the other predictions, as it is only accurate to NLO in QCD for inclusive ggF production.
All generators predict cross sections that are lower than the LHCXSWG calculation.
 The observed fiducial cross sections in the $2e2\mu$ and $2\mu2e$ final states are higher than the prediction, which leads to an overall larger observed cross section. The combined fiducial cross section and the LHCXSWG prediction agree well, only differing by 1.3 standard deviations. The $p$-values, calculated as described in Section~\ref{sec:unfolding}, are also shown in Table~\ref{tab:FiducialResults}. They indicate good compatibility with the LHCXSWG predictions.

\renewcommand{\arraystretch}{1.2}
\begin{table*}
\centering
\caption{The fiducial and total cross sections of Higgs boson production measured in the 4$\ell$ final state.  The fiducial cross sections are given separately for each decay channel, and for same- and opposite-flavour decays. The inclusive fiducial cross section is measured as the sum of all channels, as well as by combining the per-channel measurements assuming SM $\zzstar \rightarrow 4\ell$ branching ratios.
The LHCXSWG prediction is accurate to N3LO in QCD for the ggF process. For the fiducial cross-section predictions, the LHCXSWG cross sections are multiplied by the acceptances determined using the NNLOPS sample for ggF and the samples discussed in Section~\ref{sec:samples} for the other production modes.
The $p$-values indicating the compatibility of the measurement and the SM prediction are shown as well. They do not include the systematic uncertainty in the theoretical predictions.  \label{tab:FiducialResults} }
\vspace{0.2cm}
\begin{tabular}{ lcllS[table-format=1.3,table-figures-uncertainty=1]r }
\hline
\hline
Cross section [fb] & \multicolumn{3}{c}{Data ($\pm$ (stat) $\pm$ (sys)} )  & {LHCXSWG prediction} & $p$-value [\%]\\
\hline
$\sigma_{4\mu}$ & 0.92 & $^{+0.25}_{-0.23}$ & $^{+0.07}_{-0.05}$ & 0.880 \pm 0.039 & 88 \\
$\sigma_{4e}$ & 0.67 & $^{+0.28}_{-0.23}$ & $^{+0.08}_{-0.06}$ & 0.688 \pm 0.031 & 96\\
$\sigma_{2\mu2e}$ & 0.84 & $^{+0.28}_{-0.24}$ & $^{+0.09}_{-0.06}$ & 0.625 \pm 0.028 & 39\\
$\sigma_{2e2\mu}$  & 1.18 & $^{+0.30}_{-0.26}$ & $^{+0.07}_{-0.05}$ & 0.717 \pm 0.032 & 7\\
\hline
$\sigma_{4\mu+4e}$  & 1.59 & $^{+0.37}_{-0.33}$ & $^{+0.12}_{-0.10}$ & 1.57 \pm 0.07 & 65\\
$\sigma_{2\mu2e+2e2\mu}$  & 2.02 & $^{+0.40}_{-0.36}$ & $^{+0.14}_{-0.11}$  & 1.34 \pm 0.06 & 6 \\
\hline
$\sigma_\mathrm{sum}$  & 3.61 & $\pm$ 0.50 & $^{+0.26}_{-0.21}$ & 2.91 \pm 0.13 & 19 \\[0.1cm]
$\sigma_\mathrm{comb}$  & 3.62 & $\pm$ 0.50 & $^{+0.25}_{-0.20}$ & 2.91 \pm 0.13  & 18 \\
\hline
$\sigma_\mathrm{tot}$ [pb] & 69 & $^{+10}_{-9}$ & $\pm$5 & 55.6 \pm 2.5 & 19 \\[0.1cm]
\hline
\hline
\end{tabular}
\end{table*}

The measured differential cross sections and their comparisons to SM predictions are presented in Figures~\ref{fig:unfolded}$-$\ref{fig:unfolded3}.
The data are compared to SM predictions constructed from the ggF predictions provided by \progname{NNLOPS}, \mgfxfx, and, for \ptH\ and \yH, by \progname{HRes}. All ggF samples are normalized using the LHCXSWG cross section. Predictions for all other Higgs boson production modes are normalized as discussed in Section~\ref{sec:samples}.
The shaded bands on the expected cross sections indicate the PDF and scale uncertainties. The PDF inputs used for each prediction are varied according to the eigenvectors of each PDF set. The renormalization and factorization scales are varied by factors of 2.0 and 0.5. 
The figures include the $p$-values quantifying the compatibility of the measurement and the SM predictions.

\begin{figure}[!htbp]
\centering
\includegraphics[width=0.7\linewidth]{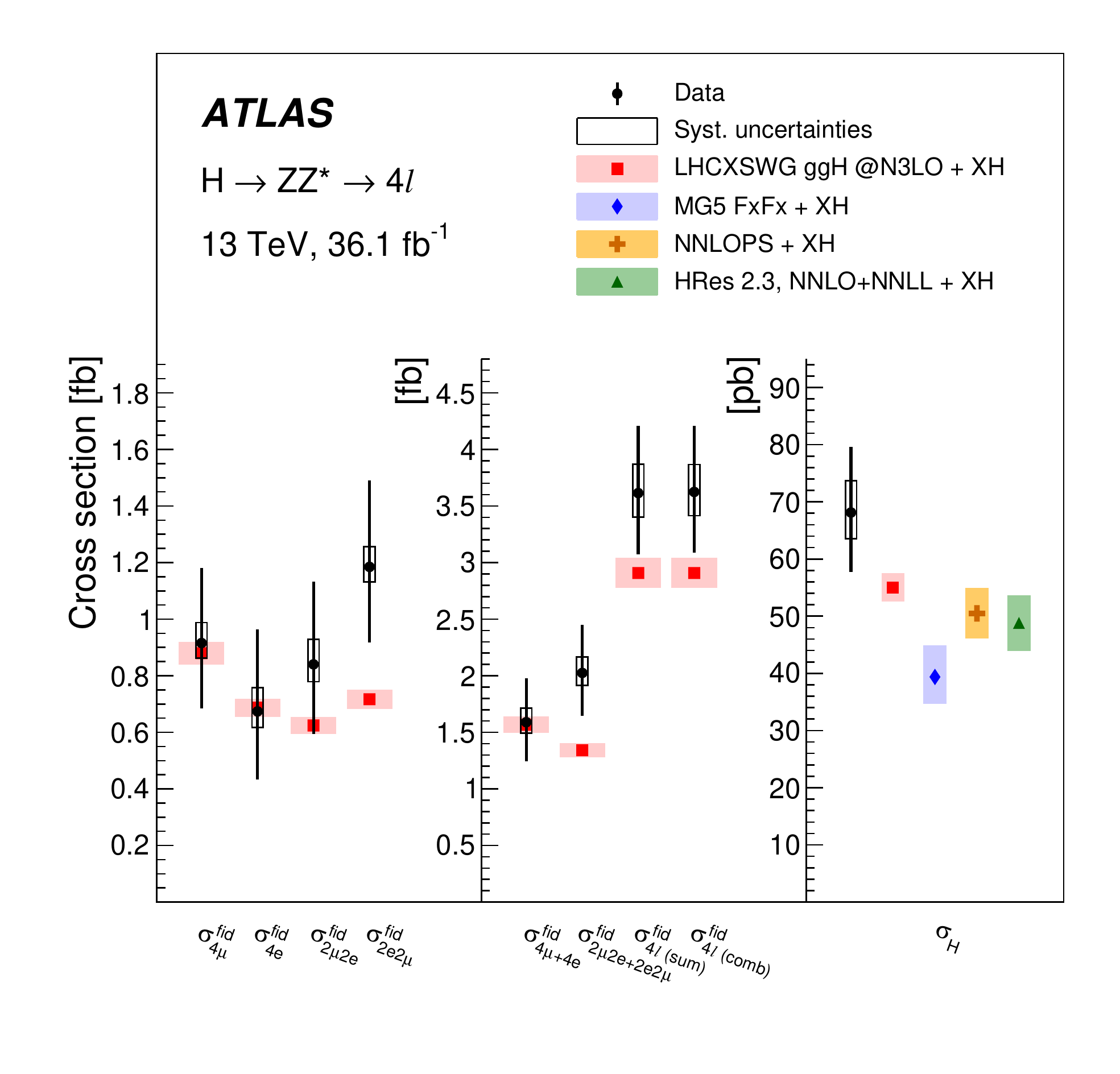}
\caption{The fiducial cross sections (left two panels) and total cross section (right panel) of Higgs boson production measured in the 4$\ell$ final state. The fiducial cross sections are shown separately for each decay channel, and for same- and opposite-flavour decays. The inclusive fiducial cross section is measured as the sum of all channels, as well as by combining the per-channel measurements assuming SM $\zzstar \rightarrow 4\ell$ branching ratios. The LHCXSWG prediction is accurate to N3LO in QCD for the ggF process. For the fiducial cross-section predictions, the LHCXSWG cross sections are multiplied by the acceptances determined using the NNLOPS sample for ggF and the samples discussed in Section~\ref{sec:samples} for the other production modes.
For the total cross section, the cross-section predictions by the generators \progname{NNLOPS}, \progname{HRes}, and \mgfxfx\ are also shown.
The cross sections for all other Higgs boson production modes $XH$ are added. The error bars on the data points show the total uncertainties, while the systematic uncertainties are indicated by the boxes.
The shaded bands around the theoretical predictions indicate the PDF and scale uncertainties. \label{fig:unfolded_fid}}
\end{figure}

\begin{figure}[!htbp]
\centering
\subfigure[\label{fig:pt}]{\includegraphics[width=0.49\linewidth]{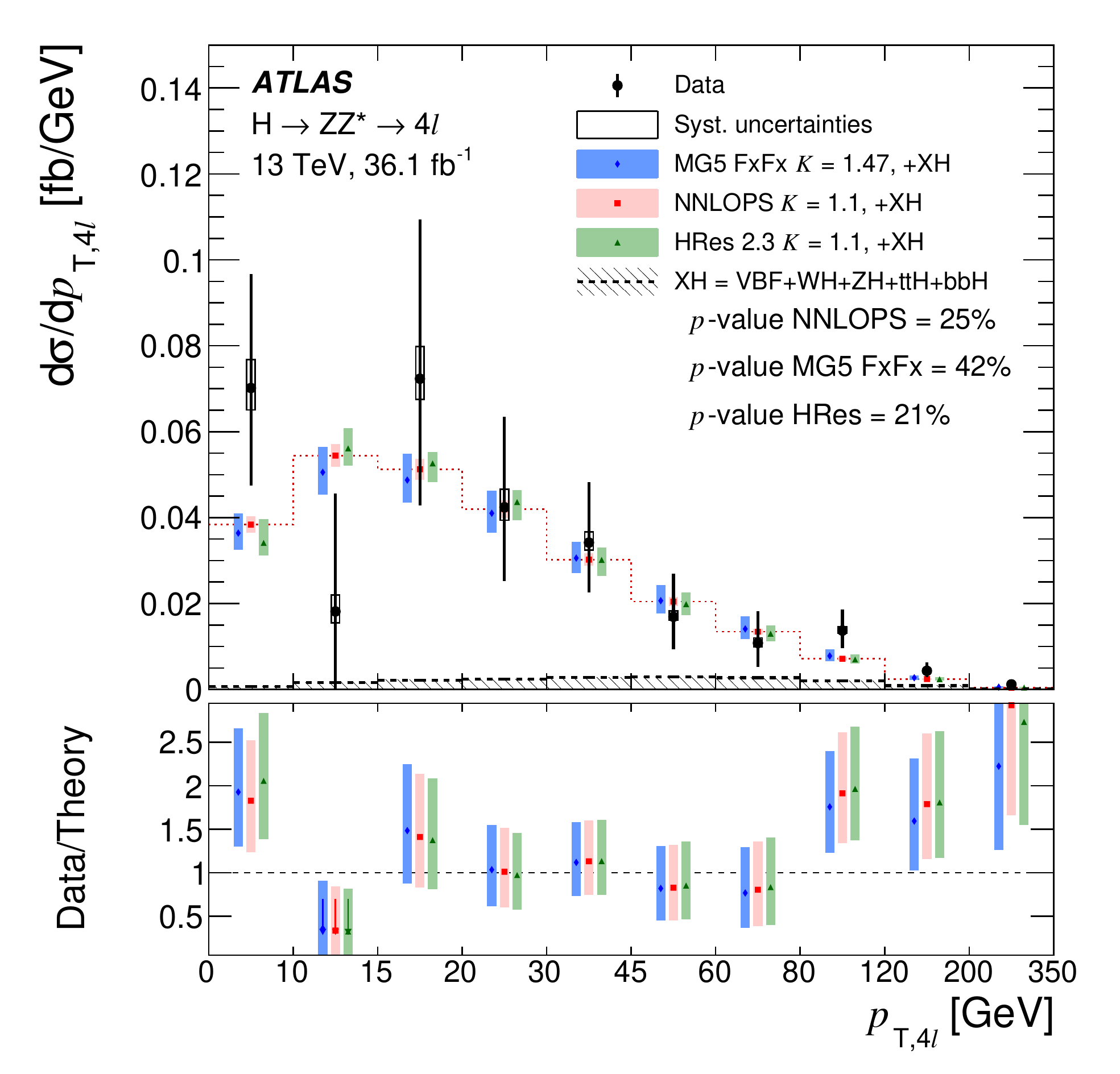}}
\subfigure[\label{fig:rap}]{\includegraphics[width=0.49\linewidth]{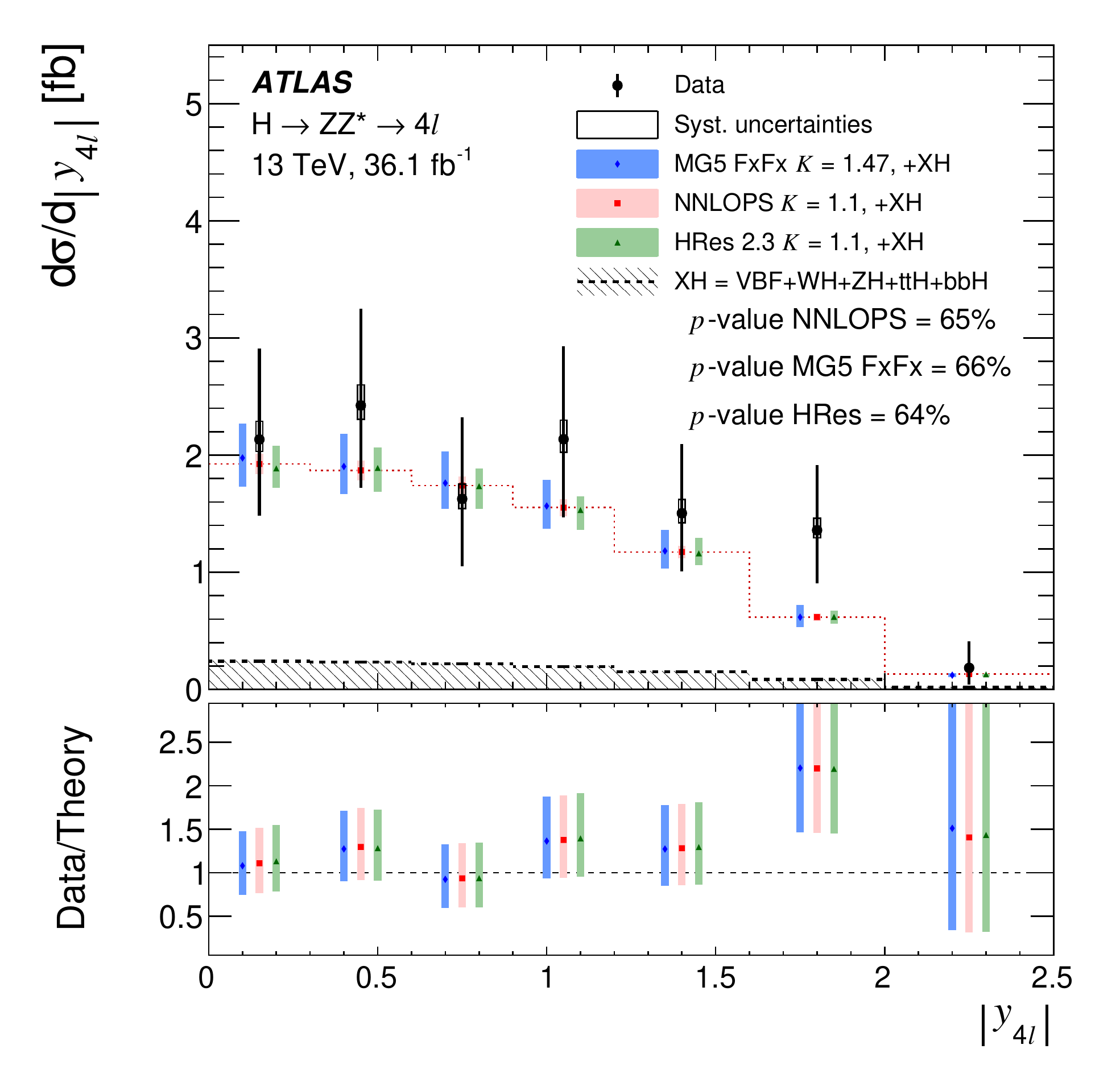}}
\subfigure[\label{fig:m34}]{\includegraphics[width=0.49\linewidth]{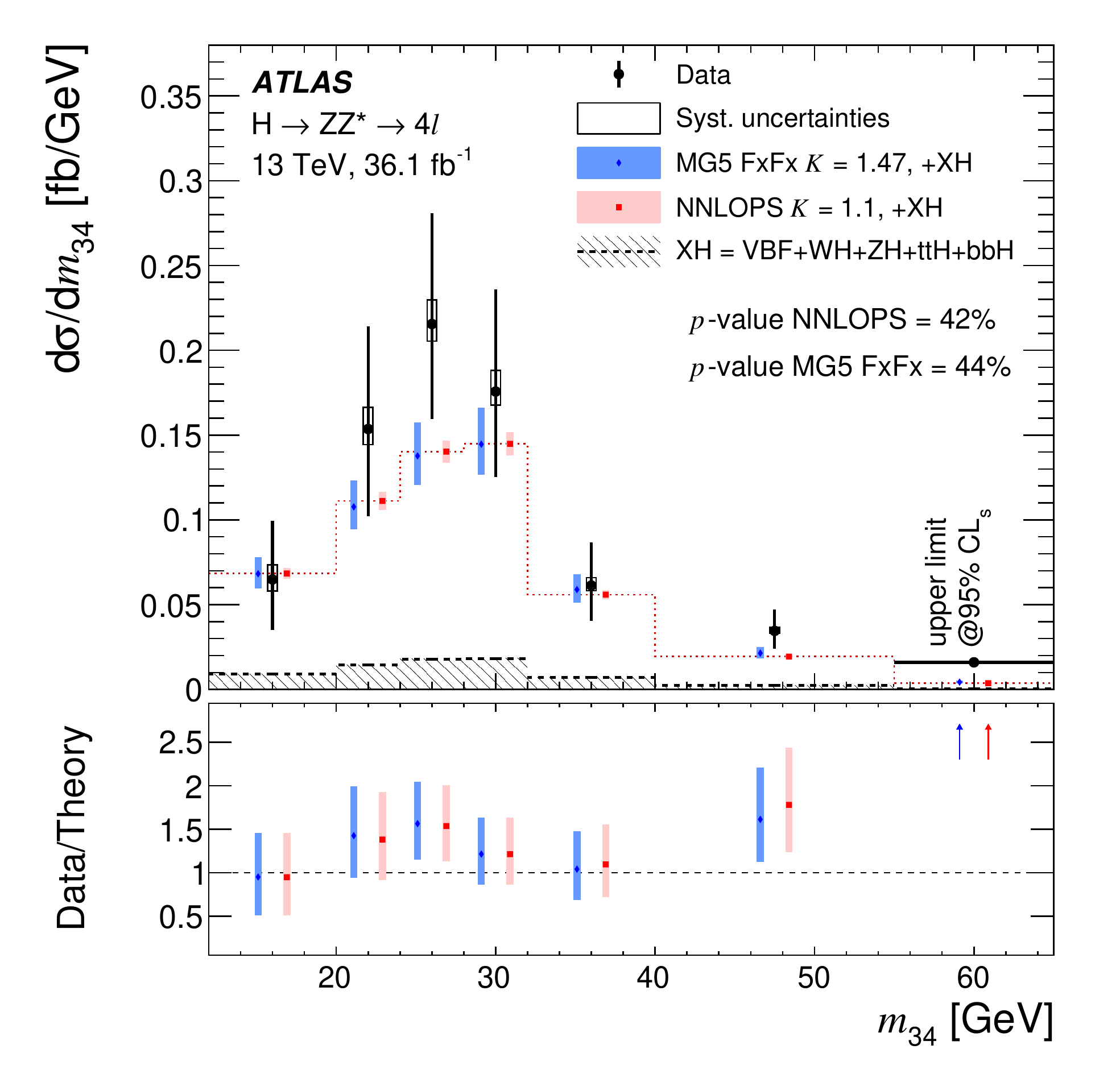}}
\subfigure[\label{fig:costheta}]{\includegraphics[width=0.49\linewidth]{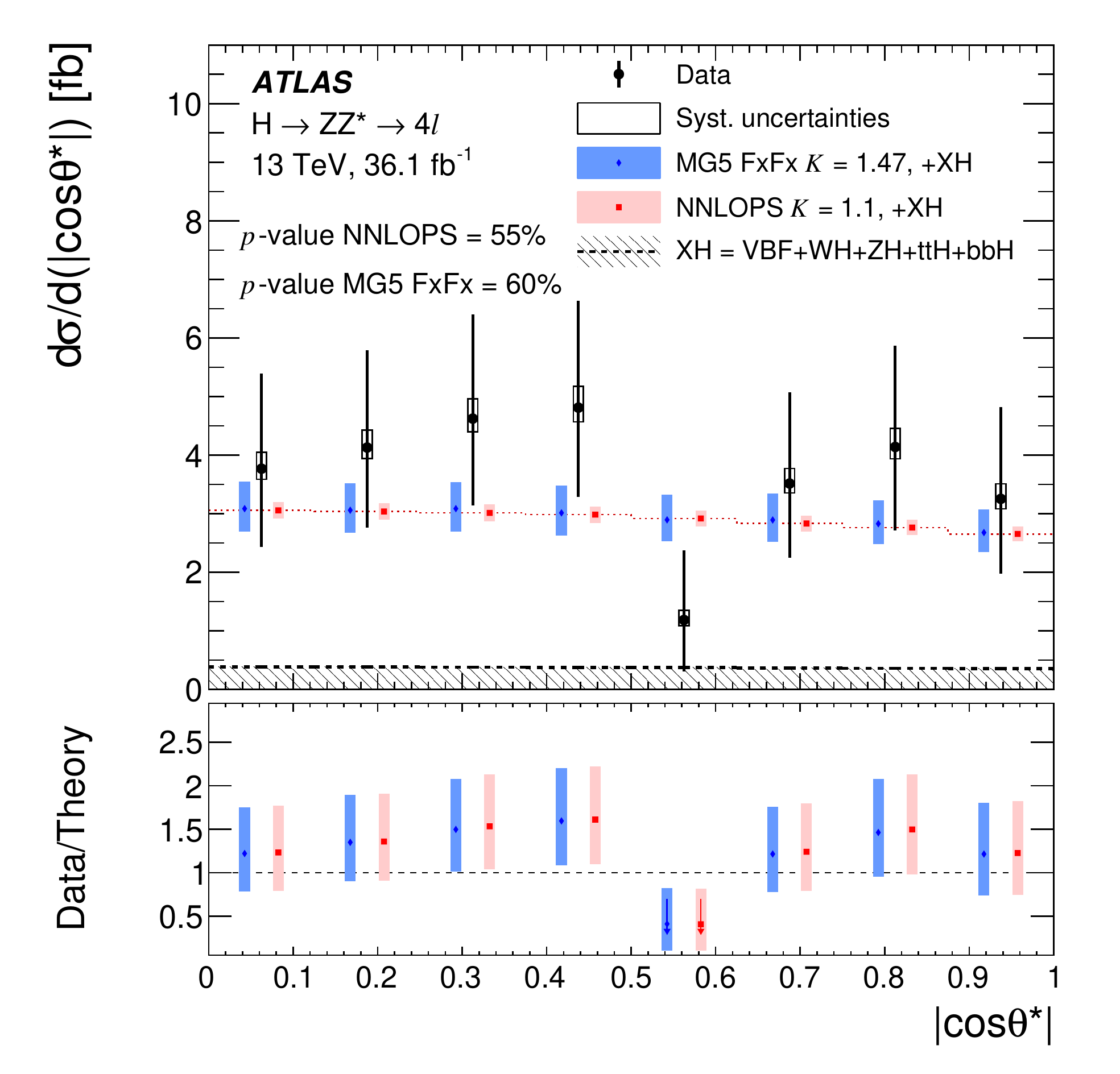}}
\caption{Differential fiducial cross sections, for (a) the transverse momentum \ptH\ of the Higgs boson, (b) the absolute value of the rapidity \yH\ of the Higgs boson, (c) the invariant mass of the subleading lepton pair \mthreefour, (d) the magnitude of the cosine of the decay angle of the leading lepton pair in the four-lepton rest frame with respect to the beam axis \costhetastar. The measured cross sections are compared to ggF predictions by \progname{NNLOPS}, \mgfxfx, and, for \ptH\ and \yH, by \progname{HRes}, all normalized to the N3LO cross section with the listed $K$-factors. Predictions for all other Higgs boson production modes $XH$ are added. The error bars on the data points show the total uncertainties, while the systematic uncertainties are indicated by the boxes. The shaded bands on the expected cross sections indicate the PDF and scale uncertainties. The $p$-values indicating the compatibility of the measurement and the SM prediction are shown as well. They do not include the systematic uncertainty in the theoretical predictions. \label{fig:unfolded}}
\end{figure}

\begin{figure}[!htbp]
\centering
\subfigure[\label{fig:njets}]{\includegraphics[width=0.49\linewidth]{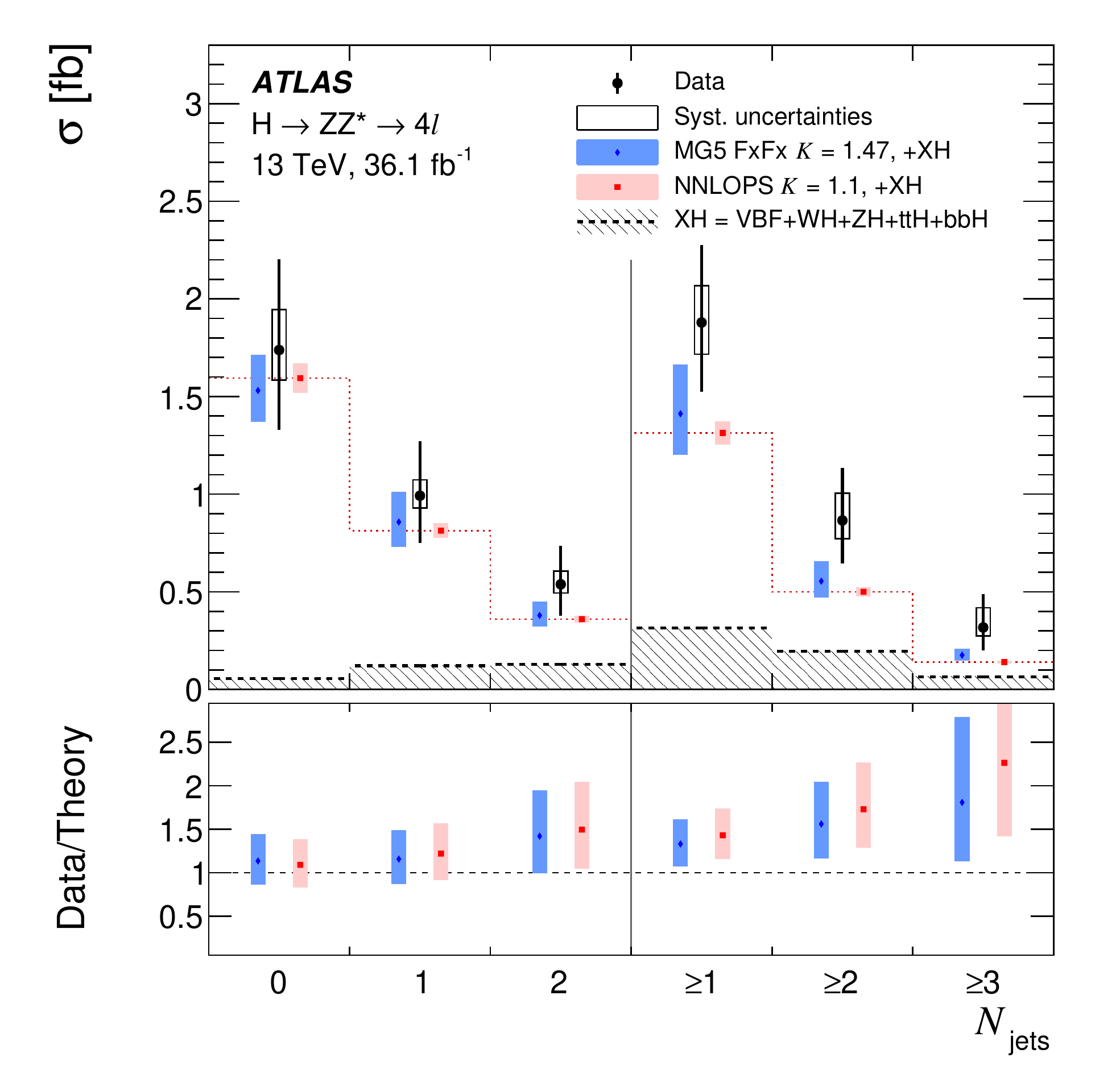}}
\subfigure[\label{fig:jetpt}]{\includegraphics[width=0.49\linewidth]{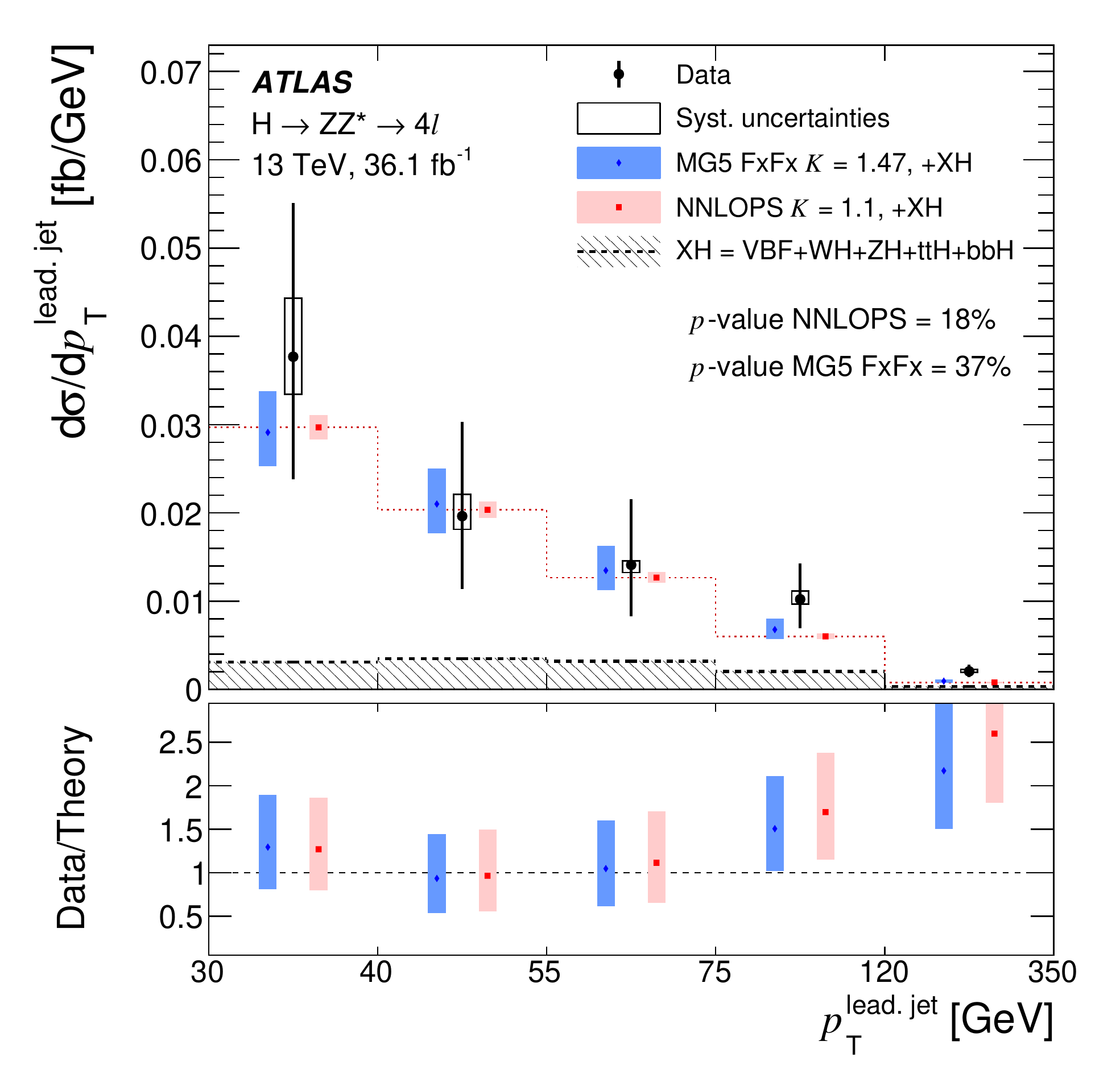}}
\subfigure[\label{fig:mjj}]{\includegraphics[width=0.49\linewidth]{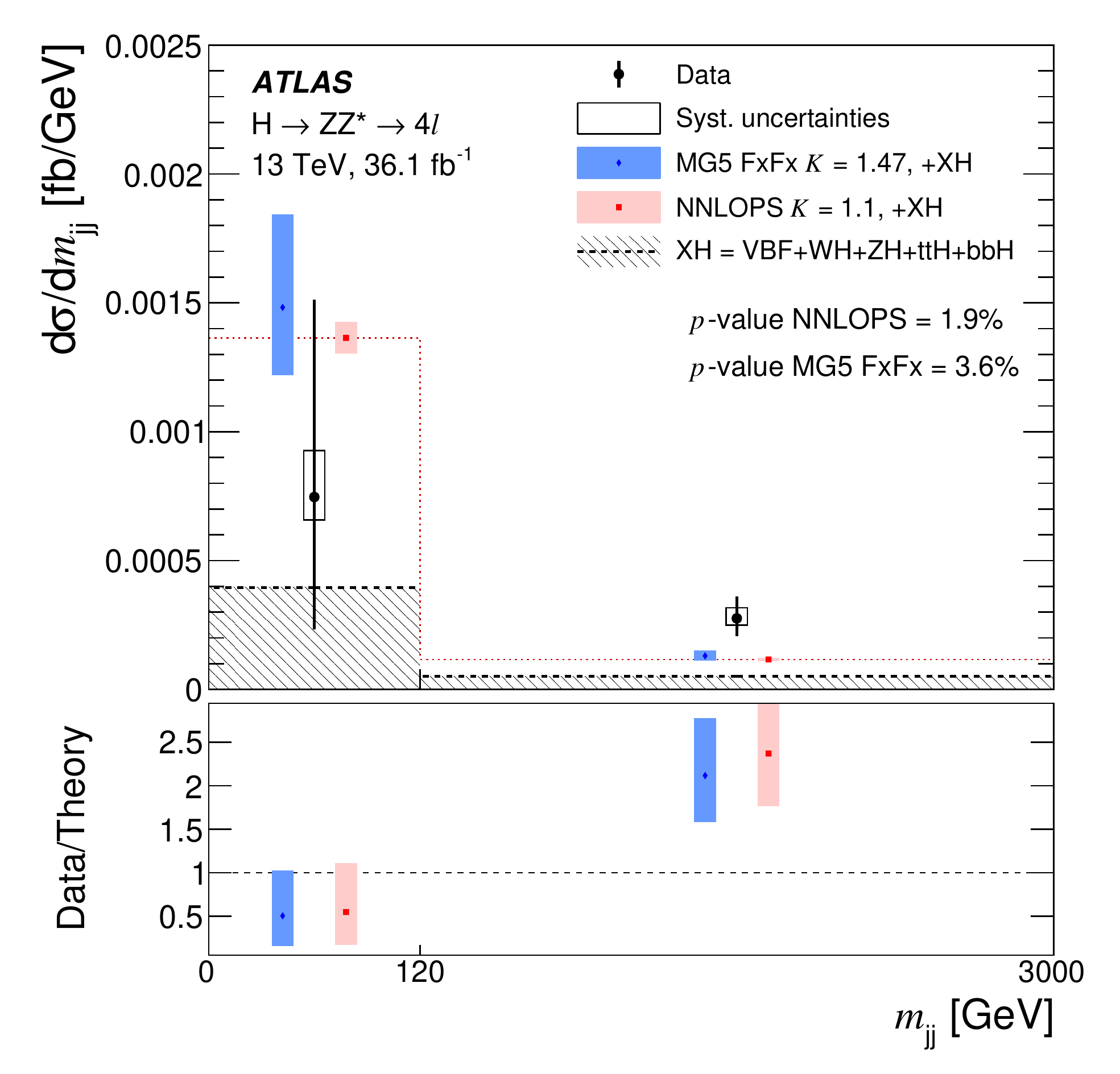}}
\subfigure[\label{fig:dphijj}]{\includegraphics[width=0.49\linewidth]{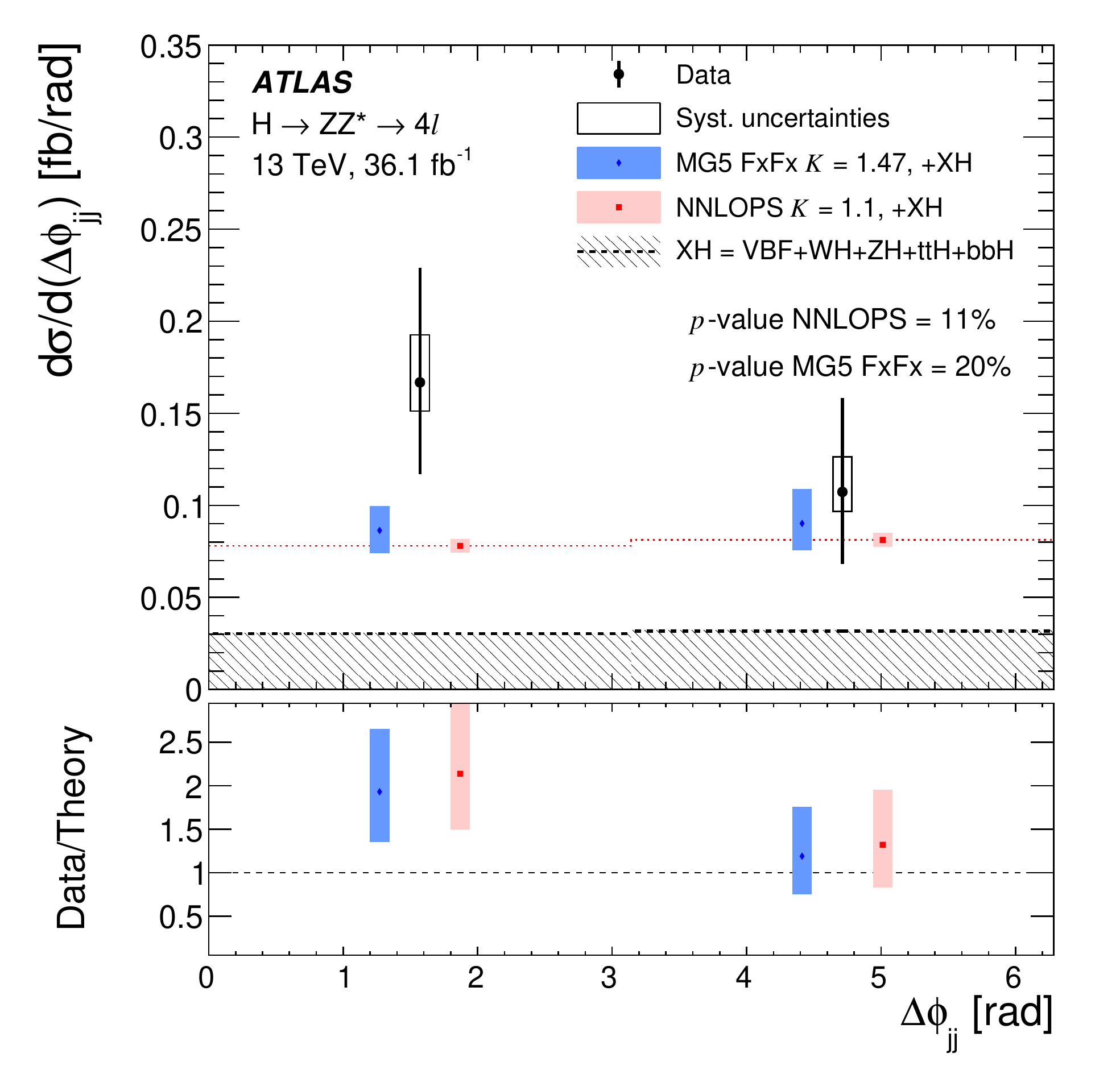}}
\caption{Differential fiducial cross sections, for (a) the number of jets \njets, (b) the transverse momentum \jetpt\ of the leading jet, (c) the invariant mass of the two leading jets \mjj, (d) the angle between the two leading jets in the transverse plane \dphijj. The measured cross sections are compared to ggF predictions by \progname{NNLOPS} and \mgfxfx, all normalized to the N3LO cross section with the listed $K$-factors. Predictions for all other Higgs boson production modes $XH$ are added.  The error bars on the data points show the total uncertainties, while the systematic uncertainties are indicated by the boxes. The shaded bands on the expected cross sections indicate the PDF and scale uncertainties. The $p$-values indicating the compatibility of the  measurement and the SM prediction are shown as well. They do not include the systematic uncertainty in the theoretical predictions.  \label{fig:unfolded2}}
\end{figure}

\begin{figure}[!htbp]
\centering
\subfigure[\label{fig:Hptvsnjets0}]{\includegraphics[width=0.49\linewidth]{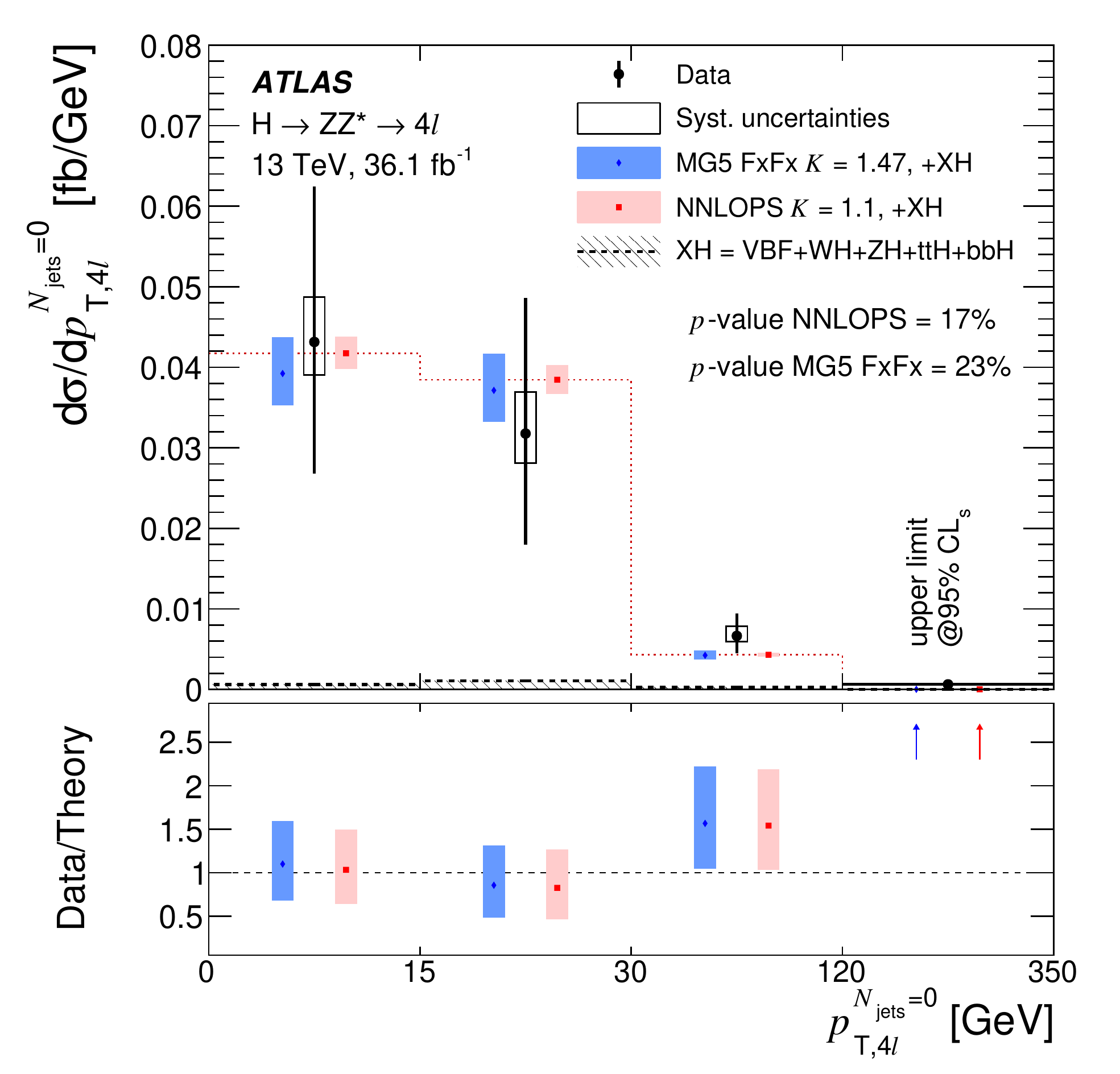}}
\subfigure[\label{fig:Hptvsnjets1}]{\includegraphics[width=0.49\linewidth]{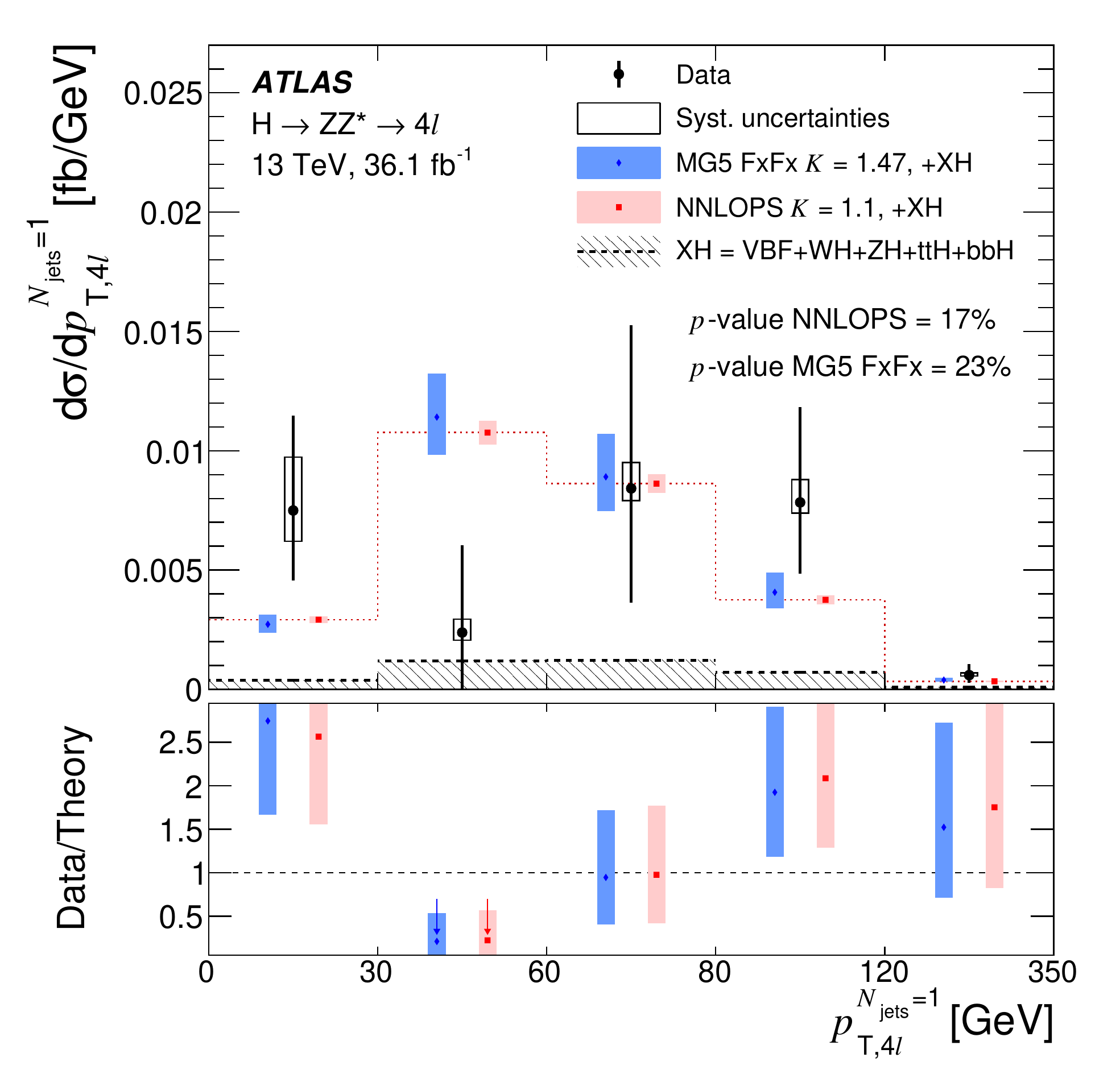}}
\subfigure[\label{fig:Hptvsnjets2}]{\includegraphics[width=0.49\linewidth]{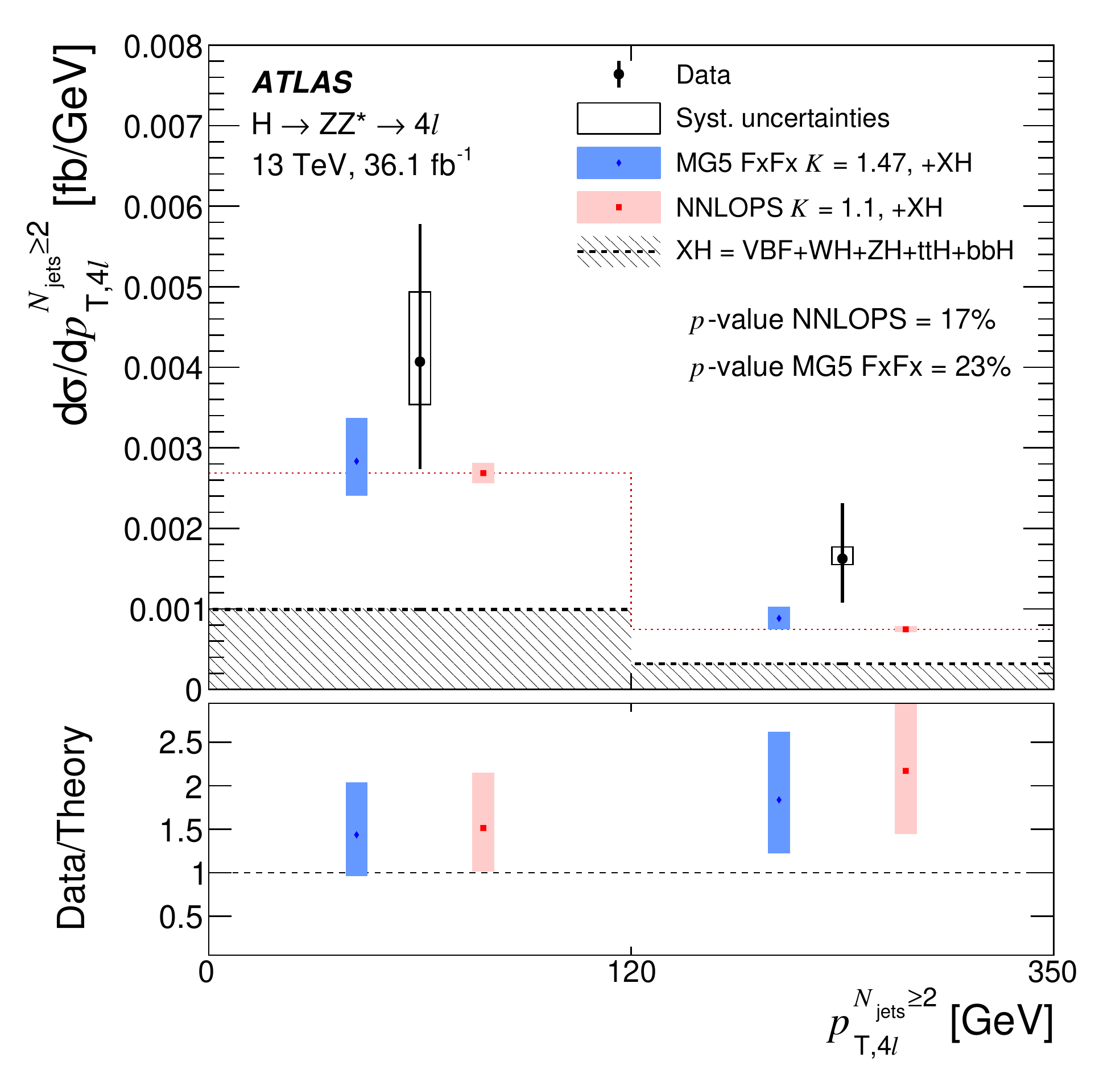}}
\subfigure[\label{fig:m12vsm34}]{\includegraphics[width=0.49\linewidth]{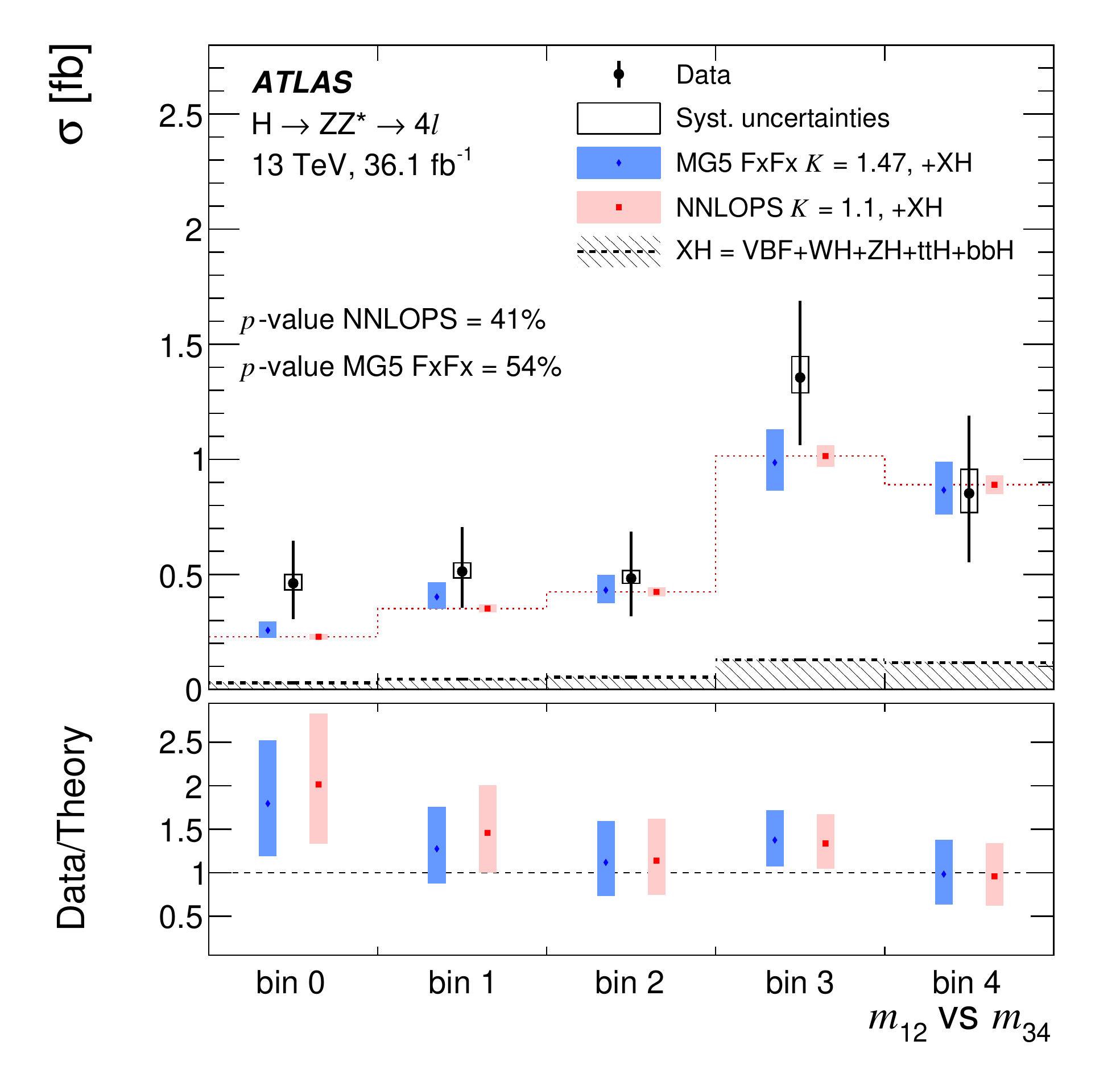}}
\caption{Figures (a)--(c) show differential fiducial cross sections of the transverse momentum \ptH\ of the Higgs boson for different jet multiplicities \njets, and (d) shows the invariant mass of the leading lepton pair  vs that of the subleading pair, \monetwo\ vs \mthreefour. The binning of \monetwo\ vs \mthreefour\ is the same as presented in Figure~\ref{fig:r_m12m34}. The measured cross sections are compared to ggF predictions by \progname{NNLOPS} and \mgfxfx, all normalized to the N3LO cross section with the listed $K$-factors. Predictions for all other Higgs boson  production modes $XH$ are added. The error bars on the data points show the total uncertainties, while the systematic uncertainties are indicated by the boxes. The shaded bands on the expected cross sections indicate the PDF and scale uncertainties. For the cross sections as a function of \ptH, the $p$-values reflect the level of agreement for the three jet bins together, treating them as a two-dimensional distribution.\label{fig:unfolded3}}
\end{figure}

The observed small excess in the measured inclusive cross section cannot be traced to a particular phase space region.
Figure~\ref{fig:unfolded} shows differential fiducial cross sections as a function of \ptH,  \yH,  \mthreefour, and \costhetastar.
The measured cross sections at high \ptH\ are slightly higher than the predictions, but the distribution is consistent with the SM predictions within the uncertainties. The observation of good agreement between data and SM prediction of the cross sections as a function of \mthreefour\ and \costhetastar\ is consistent with dedicated measurements  that have shown the Higgs boson to be a scalar particle with even parity~\cite{spin1,spin2}.

In Figure~\ref{fig:unfolded2},  the differential fiducial cross sections as a function of \njets, \jetpt, \mjj, and \dphijj\ are shown.
Agreement between data and theory is still good, but becomes a bit worse for higher jet multiplicities and higher \jetpt, similarly to what was observed in the ATLAS analyses at $\sqrt{s} = 8$~\tev~\cite{Aad:2014tca,Aad:2014lwa,Aad:2015lha}. \mgfxfx\ describes the jet multiplicities slightly better than \progname{NNLOPS}. For large values of \mjj\ and the left bin of the \dphijj\ distribution, the measured cross section is more than twice the predicted value ($\sim$2 and $\sim$1.5 standard deviations respectively).

Figure~\ref{fig:unfolded3} presents the differential fiducial cross sections as a function of \ptH\ for different jet multiplicities as well as the cross sections measured in regions of the \monetwo\ vs \mthreefour\ distribution. For the latter, the \monetwo\ vs \mthreefour\ kinematic plane is divided into five regions and projected onto a one-dimensional distribution, as shown in Figure~\ref{fig:r_m12m34}.
The split into different jet multiplicities allows one to probe perturbative QCD calculations for different production modes. The 0-jet bin is dominated by Higgs boson events produced through ggF, while the $\ge2$-jet bin is enriched with VBF events. No significant deviation from the predictions is seen, as indicated by the $p$-values which reflect the level of agreement for the three jet bins together, treating them as a two-dimensional distribution. The higher values of the measured cross sections in the $\ge2$-jet bin reflect the observations in Figure~\ref{fig:njets}.  The data and the predictions also agree well for the \monetwo\ vs \mthreefour\ distribution.

\begin{figure}[!htbp]
\centering
\subfigure[\label{fig:PO_eler}]{\includegraphics[width=0.49\linewidth]{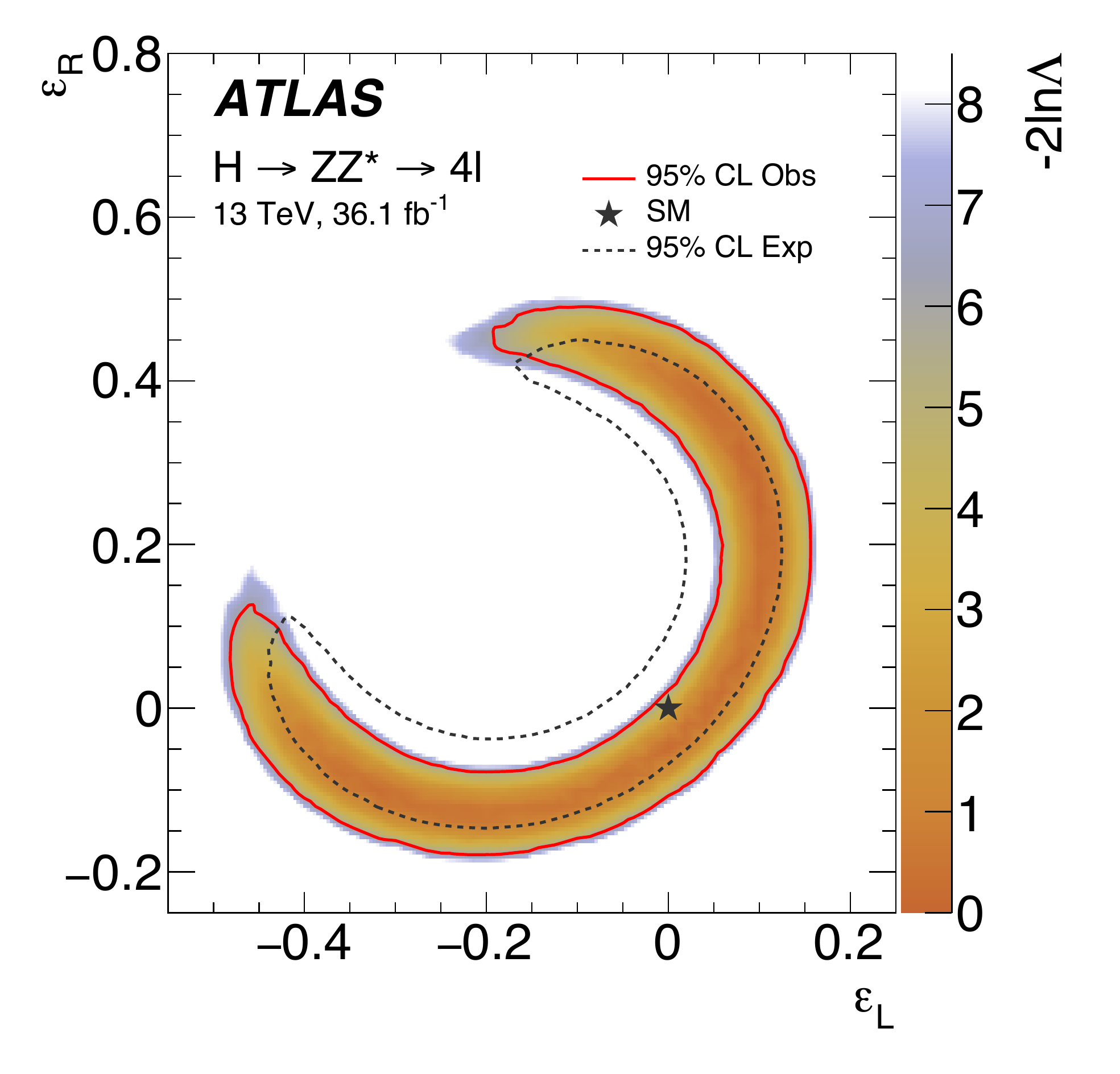}}
\subfigure[\label{fig:PO_kzzel}]{\includegraphics[width=0.49\linewidth]{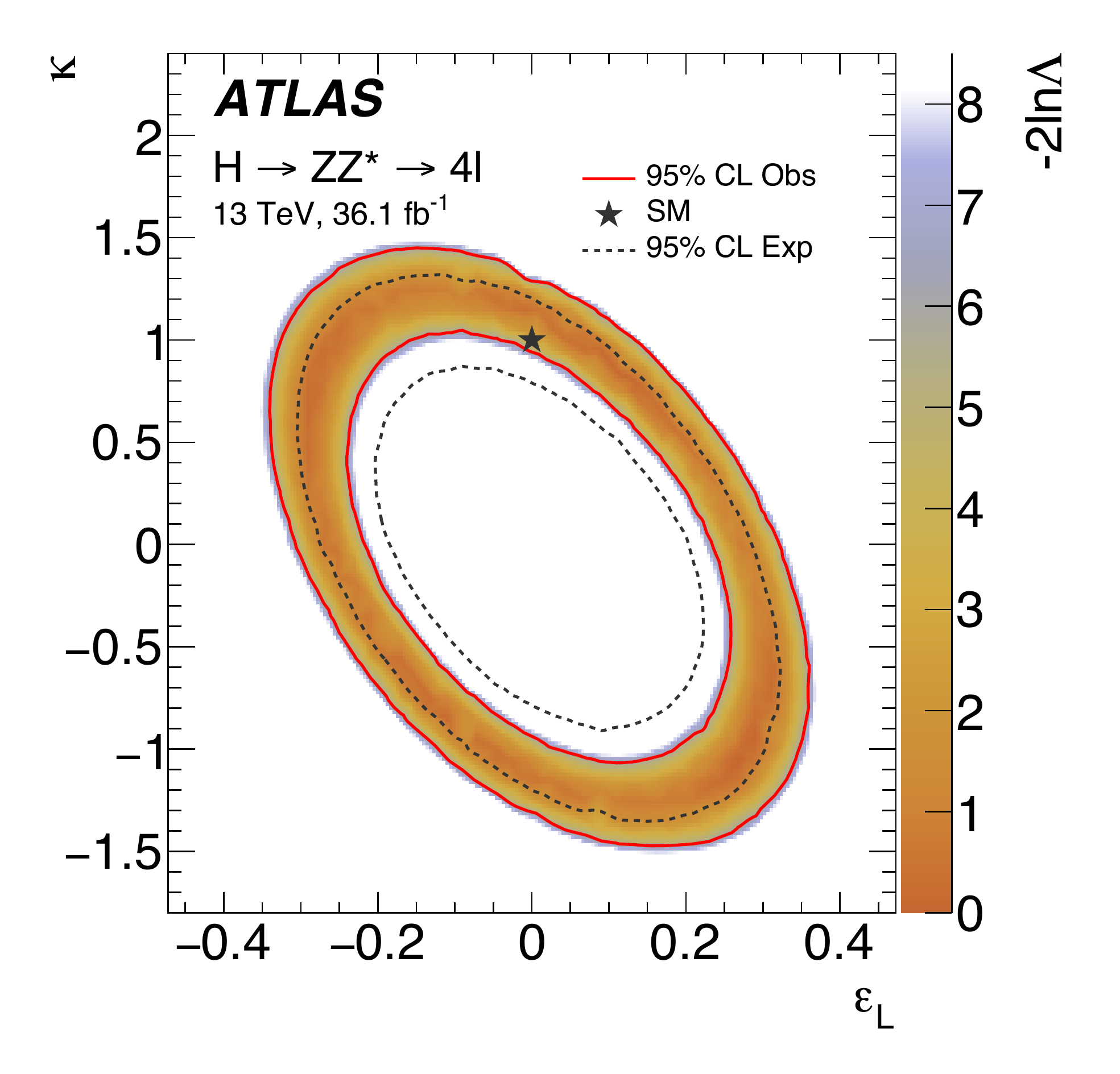}}
\caption{Limits on modified Higgs boson decays within the framework of pseudo-observables~\cite{Gonzalez-Alonso:2014eva, Gonzalez-Alonso:2015eva}. In (a), the limits are extracted in the plane of $\varepsilon_\mathrm{L}$ and $\varepsilon_\mathrm{R}$, which modify the contact terms between the Higgs boson and left- and  right-handed leptons, assuming lepton-flavour universality. In (b), the tested parameters are $\varepsilon_\mathrm{L}$ and $\kappa$. The latter modifies the coupling of the Higgs boson to $Z$ bosons. The allowed observed area at the 95\% CL is surrounded by the red solid line. This  can be compared to the SM prediction, which is indicated by the black star and the black dotted line. The coloured scale indicates the values of $-2 \ln{\Lambda}$. \label{fig:PO}}
\end{figure}

The differential fiducial cross sections can be interpreted in the context of searches for physics beyond the SM.
In the absence of significant deviations from the SM predictions, limits are set on modified Higgs boson interactions within the framework of pseudo-observables~\cite{Gonzalez-Alonso:2014eva, Gonzalez-Alonso:2015eva}.
In this paper, the couplings related to the contact interaction of the Higgs boson decay are considered, $\varepsilon_\mathrm{L}$ and  
$\varepsilon_\mathrm{R}$, which modify, in a flavour-universal way, the contact terms between the Higgs boson, the $Z$ boson, and left- or right-handed leptons.
Since the contact terms have the same Lorentz structure as the SM term, they only affect the dilepton invariant mass spectra, while the lepton angular distributions are not modified. The difference in $\chi^2$ between the measured and predicted cross sections in the \monetwo\ vs \mthreefour\ observable plane is therefore used to constrain the possible contributions from contact interactions.  It was checked with pseudo experiments that the $\chi^2$ distribution agrees with the hypothesis of two degrees of freedom.
Assuming the SM values for all but the tested parameters, limits are set on the contact-interaction coupling strength as shown in Figure~\ref{fig:PO}. Two parameter planes are considered: $\varepsilon_\mathrm{L}$ vs $\varepsilon_\mathrm{R}$, as well as $\varepsilon_\mathrm{L}$ vs $\kappa$, where $\kappa$ is the coupling of the Higgs boson to the $Z$ bosons and $\varepsilon_\mathrm{R} = 0.48 \cdot \varepsilon_\mathrm{L}$~\cite{Gonzalez-Alonso:2015eva}.
Since the addition of the contact terms changes the Higgs boson production rate, in principle limits could be set based on the inclusive Higgs boson cross sections alone. In this case, the obtained allowed area in Figure~\ref{fig:PO_eler} would be circular, but the addition of the invariant mass spectra improves the sensitivity, especially for negative $\varepsilon_\mathrm{L}$ and positive $\varepsilon_\mathrm{R}$. The addition of the shape information also improves the limit in the $\varepsilon_\mathrm{L}$ vs $\kappa$ parameter plane.
It can be seen that the expected and observed limits are slightly shifted with respect to each other, but no significant deviation is observed.

\section{Conclusion}
\label{sec:conclusions}
Measurements of the inclusive and differential fiducial cross sections of Higgs boson production in the \hfourl\ decay channel are presented. They are based on data extracted from 36.1~fb$^{-1}$ of $\sqrt{s} = 13$~\tev\ proton--proton collisions recorded by the ATLAS detector at the LHC in 2015 and 2016.
  The inclusive fiducial cross section in the $H \rightarrow ZZ^* \rightarrow 4\ell$ decay channel is measured to be \hbox{\mxsec\mxsecErr~fb}, in agreement with the Standard Model prediction of \hbox{\mxsecExp\mxsecExpErr~fb}. The inclusive fiducial cross section is also extrapolated to the total phase space which includes all Standard Model Higgs boson decays. Several differential fiducial cross sections are measured for observables sensitive to the Higgs boson production and decay, including kinematic distributions of the jets produced together with the Higgs boson. Good agreement is found between the data and the predictions of the Standard Model. 
The extracted cross-section distributions are used to constrain anomalous Higgs boson interactions with Standard Model particles using the pseudo-observable framework.

\section*{Acknowledgements}


We thank CERN for the very successful operation of the LHC, as well as the
support staff from our institutions without whom ATLAS could not be
operated efficiently.

We acknowledge the support of ANPCyT, Argentina; YerPhI, Armenia; ARC, Australia; BMWFW and FWF, Austria; ANAS, Azerbaijan; SSTC, Belarus; CNPq and FAPESP, Brazil; NSERC, NRC and CFI, Canada; CERN; CONICYT, Chile; CAS, MOST and NSFC, China; COLCIENCIAS, Colombia; MSMT CR, MPO CR and VSC CR, Czech Republic; DNRF and DNSRC, Denmark; IN2P3-CNRS, CEA-DSM/IRFU, France; SRNSF, Georgia; BMBF, HGF, and MPG, Germany; GSRT, Greece; RGC, Hong Kong SAR, China; ISF, I-CORE and Benoziyo Center, Israel; INFN, Italy; MEXT and JSPS, Japan; CNRST, Morocco; NWO, Netherlands; RCN, Norway; MNiSW and NCN, Poland; FCT, Portugal; MNE/IFA, Romania; MES of Russia and NRC KI, Russian Federation; JINR; MESTD, Serbia; MSSR, Slovakia; ARRS and MIZ\v{S}, Slovenia; DST/NRF, South Africa; MINECO, Spain; SRC and Wallenberg Foundation, Sweden; SERI, SNSF and Cantons of Bern and Geneva, Switzerland; MOST, Taiwan; TAEK, Turkey; STFC, United Kingdom; DOE and NSF, United States of America. In addition, individual groups and members have received support from BCKDF, the Canada Council, CANARIE, CRC, Compute Canada, FQRNT, and the Ontario Innovation Trust, Canada; EPLANET, ERC, ERDF, FP7, Horizon 2020 and Marie Sk{\l}odowska-Curie Actions, European Union; Investissements d'Avenir Labex and Idex, ANR, R{\'e}gion Auvergne and Fondation Partager le Savoir, France; DFG and AvH Foundation, Germany; Herakleitos, Thales and Aristeia programmes co-financed by EU-ESF and the Greek NSRF; BSF, GIF and Minerva, Israel; BRF, Norway; CERCA Programme Generalitat de Catalunya, Generalitat Valenciana, Spain; the Royal Society and Leverhulme Trust, United Kingdom.

The crucial computing support from all WLCG partners is acknowledged gratefully, in particular from CERN, the ATLAS Tier-1 facilities at TRIUMF (Canada), NDGF (Denmark, Norway, Sweden), CC-IN2P3 (France), KIT/GridKA (Germany), INFN-CNAF (Italy), NL-T1 (Netherlands), PIC (Spain), ASGC (Taiwan), RAL (UK) and BNL (USA), the Tier-2 facilities worldwide and large non-WLCG resource providers. Major contributors of computing resources are listed in Ref.~\cite{ATL-GEN-PUB-2016-002}.

\clearpage

\printbibliography

\newpage 
\begin{flushleft}
{\Large The ATLAS Collaboration}

\bigskip

M.~Aaboud$^\textrm{\scriptsize 137d}$,
G.~Aad$^\textrm{\scriptsize 88}$,
B.~Abbott$^\textrm{\scriptsize 115}$,
O.~Abdinov$^\textrm{\scriptsize 12}$$^{,*}$,
B.~Abeloos$^\textrm{\scriptsize 119}$,
S.H.~Abidi$^\textrm{\scriptsize 161}$,
O.S.~AbouZeid$^\textrm{\scriptsize 139}$,
N.L.~Abraham$^\textrm{\scriptsize 151}$,
H.~Abramowicz$^\textrm{\scriptsize 155}$,
H.~Abreu$^\textrm{\scriptsize 154}$
% ...
\bigskip
\\
% ...
$^{12}$ Institute of Physics, Azerbaijan Academy of Sciences, Baku, Azerbaijan\\
% ...
$^{88}$ CPPM, Aix-Marseille Universit{\'e} and CNRS/IN2P3, Marseille, France\\
% ...
$^{115}$ Homer L. Dodge Department of Physics and Astronomy, University of Oklahoma, Norman OK, United States of America\\
% ...
$^{119}$ LAL, Univ. Paris-Sud, CNRS/IN2P3, Universit{\'e} Paris-Saclay, Orsay, France\\
% ...
$^{137}$ $^{(a)}$ Facult{\'e} des Sciences Ain Chock, R{\'e}seau Universitaire de Physique des Hautes Energies - Universit{\'e} Hassan II, Casablanca; $^{(b)}$ Centre National de l'Energie des Sciences Techniques Nucleaires, Rabat; $^{(c)}$ Facult{\'e} des Sciences Semlalia, Universit{\'e} Cadi Ayyad, LPHEA-Marrakech; $^{(d)}$ Facult{\'e} des Sciences, Universit{\'e} Mohamed Premier and LPTPM, Oujda; $^{(e)}$ Facult{\'e} des sciences, Universit{\'e} Mohammed V, Rabat, Morocco\\
% ...
$^{139}$ Santa Cruz Institute for Particle Physics, University of California Santa Cruz, Santa Cruz CA, United States of America\\
% ...
$^{151}$ Department of Physics and Astronomy, University of Sussex, Brighton, United Kingdom\\
% ...
$^{154}$ Department of Physics, Technion: Israel Institute of Technology, Haifa, Israel\\
$^{155}$ Raymond and Beverly Sackler School of Physics and Astronomy, Tel Aviv University, Tel Aviv, Israel\\
% ...
$^{161}$ Department of Physics, University of Toronto, Toronto ON, Canada\\
% ...
$^{a}$ Also at Department of Physics, King's College London, London, United Kingdom\\
$^{b}$ Also at Institute of Physics, Azerbaijan Academy of Sciences, Baku, Azerbaijan\\
$^{c}$ Also at Novosibirsk State University, Novosibirsk, Russia\\
$^{d}$ Also at TRIUMF, Vancouver BC, Canada\\
$^{e}$ Also at Department of Physics {\&} Astronomy, University of Louisville, Louisville, KY, United States of America\\
% ...
$^{*}$ Deceased
\end{flushleft}


\end{document}